\documentclass[a4paper,11pt]{article}
\usepackage{setspace}
\usepackage[left = 2cm, right=2cm, top=3cm, bottom=3cm]{geometry}
\usepackage[pdftex]{graphicx}
\usepackage{cite}
\usepackage{bm}\usepackage{pdfpages}
\usepackage{float}
\usepackage{amsmath,amssymb,latexsym}
\usepackage{color}
\usepackage[T1]{fontenc}
\usepackage{wasysym}
\usepackage{siunitx}
\usepackage{setspace}
\usepackage{relsize}
\usepackage{tikz}
\usepackage{lineno, blindtext}
\usepackage{authblk}
\usepackage[symbol]{footmisc}
\usepackage{lineno}
\usepackage[pdfencoding=auto, psdextra]{hyperref}

\usepackage{mismath}
\begin{document}

	\renewcommand{\figurename}{Fig.}
	
	\title{\color{blue}\textbf{Investigating the {roles} of hydrophobicity and electrostatics {in} the particle-scale dynamics {and rheology} of dense microgel suspensions}}
	\author[1]{Sayantan Chanda}
	\affil[1]{\textit{Soft Condensed Matter Group, Raman Research Institute, C. V. Raman Avenue, Sadashivanagar, Bangalore 560 080, INDIA}}
	\author[1]{Chandeshwar Misra}
	\author[1,*]{Ranjini Bandyopadhyay}
	
	\footnotetext[1]{Corresponding Author: Ranjini Bandyopadhyay; Email: ranjini@rri.res.in}
	\maketitle
	\begin{abstract}
Colloidal microgel particles such as poly(N-isopropylacrylamide) (PNIPAM) {shrink reversibly in an aqueous medium due to the expulsion of water} at a volume phase transition
temperature, VPTT {$\sim $33$^ {\circ}$C}. 
Romeo {{et. al}} [Adv. Mater. 2010, {\textbf{22}}, 3441–3445] had previously shown that dense aqueous PNIPAM suspensions transformed from
one viscoelastic solid-like phase to another {when suspension temperature was increased}, with an intermediate viscoelastic liquid-like phase
near the VPTT. {They attributed this observation} to a change in the inter-particle interaction from hydrophilic to hydrophobic. Here, we show
using a combination of experimental techniques that particle hydrophobicity can become significant even below the VPTT. We achieve this by incorporating {dissociating} additives such as sodium chloride and potassium chloride, or {non-dissociating} additives such as sucrose, into the aqueous medium. Above the VPTT, we observe that
suspension rigidity is the highest in the presence of salts because of the combined effects of electrostatic and hydrophobic attractions. In the
presence of non-dissociating sucrose, in contrast, the inter-microgel interaction remains hydrophobic across the VPTT. Such easy tunability of interactions by
incorporating commonly available chemicals into the suspension medium opens up new avenues for the synthesis of novel metamaterials. 
  \end{abstract}

\section{Introduction:}
Colloidal microgel particles, comprising chemically cross-linked three-dimensional polymer networks, are compressible, deformable, and { swell and shrink in response} to external stimuli such as temperature, pressure, pH, ionic strength, solvent medium, etc.~\cite{snowden1996colloidal,lyon2004microgel,lyon2012polymer,fernandez2003osmotic,buratti2025copolymer,cho2009tunable,vlassopoulos2014tunable}. Poly(N-isopropylacrylamide) or PNIPAM is one such thermo-responsive microgel.
Aqueous suspensions of PNIPAM particles undergo a reversible volume phase
transition at a volume phase transition temperature, VPTT, of approximately 33$^{\circ}$C~\cite{heskinS1968solution,franco2025soft}. 
The interaction between microgel particles is governed by a combination of steric hindrance, screened  Coulombic repulsion, hydrophobicity, and van der Waals interaction~\cite{minami2019rheological, rasmusson2004flocculation}. Below the VPTT, solvent-polymer interaction dominates and PNIPAM microgels behave as hydrophilic particles. They {swell by absorbing} water and interact mainly \textit{\textit{via}} steric repulsion~\cite{minami2016rheological, minami2019rheological}. The particles become hydrophobic when suspension temperature is raised above the VPTT. They shrink in size by expelling water in a process that leads to a loss of steric repulsion. Charges on the microgel surfaces, arising from charged initiators and polar cross-linkers used during synthesis, can electrostatically stabilize the microgels above the VPTT~\cite{misra2024effect,town2019understanding}. 

The thermo-responsive nature of PNIPAM particles has been exploited to study a wide variety of viscoelastic phases~\cite{minami2016rheological, senff1999temperature,romeo2010temperature,wang2014revisit,chaudhary2021linear,vialetto2024effect,misra2020influence,ghosh2019linear,franco2021glass}. Rheological experiments have previously reported that dense PNIPAM suspensions transform from a repulsive glassy phase to an attractive gel phase {with increase in suspension temperature, with an intermediate} viscoelastic liquid-like phase near the VPTT~\cite{romeo2010temperature,wang2014revisit}. More recently, Misra et al.~\cite{misra2024effect} have demonstrated that the phase behavior of dense PNIPAM suspensions is influenced by the stiffness of individual microgels. Notably, an intermediate liquid-like phase at the VPTT was observed only when the PNIPAM microgels were highly soft and deformable. In contrast, the rheological response of suspensions prepared with stiffer particles remained viscoelastic solid-like even near the VPTT. 

{The incorporation of additives can significantly influence the viscoelastic behavior of colloidal suspensions by modifying inter-particle and solvent-particle interactions}~\cite{sagle2009investigating,misra2021influence,jeldres2014impact,ranganathan2017effects,costa2015adjusting,minami2020criteria}.   
Previous studies {of aqueous PNIPAM suspensions} have shown that the inclusion of dissociating salts like {sodium chloride, potassium chloride}, etc., or non-dissociating osmolytes like glucose or sucrose can enhance the hydrophobicity of PNIPAM particles~\cite{paz2004interaction,narang2017unexplored,zhang2005specific}. This results in the lowering of the VPTTs of PNIPAM suspensions~\cite{narang2017unexplored, costa2015adjusting}. Additionally, salts can also suppress electrostatic repulsion among PNIPAM particles~\cite{rasmusson2004flocculation,daly2000temperature,minami2020criteria}, leading to large-scale particle flocculation. 
While the {roles} of hydrophobicity and electrostatics on PNIPAM particles {and their suspensions have} been studied separately~\cite{narang2017unexplored, zhang2005specific, rasmusson2004flocculation, minami2020criteria}, their combined effect on the bulk viscoelastic properties of dense PNIPAM suspensions remains largely unexplored. 

In the present study, we employed an array of experimental techniques to study the impact of hydrophobic and electrostatic interactions on the bulk rheological properties of dense aqueous suspensions of PNIPAM particles. Controlled and systematic variations in particle hydrophobicity and inter-particle electrostatic interactions were achieved \textit{\textit{via}} the {incorporation} of dissociating ({sodium chloride} and {potassium chloride}) and non-dissociating (sucrose) additives to the suspension medium. Dynamic light scattering (DLS) experiments showed a reduction in particle sizes upon additive inclusion at a temperature below the VPTT, while temperature-sweep oscillatory rheology displayed a loss in suspension rigidity. 
Simultaneously, our cryogenic field emission scanning electron microscopy (cryo-FESEM) data displayed more porous suspension structures in the presence of additives and agreed well with the observed reduction in mechanical moduli values. We argue that hydrophobicity-driven particle shrinkage below the VPTT in the presence of suitable additive drives microgel association and the physicochemical properties of dense PNIPAM suspensions.

Near and above the VPTT, dense PNIPAM suspensions displayed colloidal gel-like behavior. Rheological data acquired above the VPTT suggests that the suspensions containing dissociating additives formed stronger and more rigid gels. This observation was again supported by the cryo-FESEM images, which showed denser space-spanning networks with thicker gel strands in samples prepared with dissociating salts. Based on our findings from Fourier transform infrared spectroscopy (FTIR) and zeta potential experiments, we conclude that the combined effect of hydrophobic and electrostatic attraction drives stronger gelation above the VPTT in the presence of dissociating salts. In the presence of non-dissociating sucrose, however, inter-particle attraction above the VPTT is driven solely by suspension hydrophobicity, leading to the formation of comparatively weaker gels. 

In summary, while the contribution of hydrophobicity is the most important factor in determining PNIPAM microgel suspension properties in the presence of dissociating and non-dissociating additives below the VPTT, a combined effect of hydrophobicity and electrostatics dominates near and above the VPTT. The physicochemical properties of PNIPAM suspensions can, therefore, be controlled by carefully tuning the interplay between hydrophobic and electrostatic interactions. {We conclude that the} incorporation of dissociating and non-dissociating additives to aqueous PNIPAM suspensions is {an effective} strategy to fine-tune inter-particle interactions, particle self-assembly, and suspension viscoelasticity.

\section{\label{em}Materials and methods}
\subsection{Synthesis and sample preparation:}
{
N-isopropylacrylamide (NIPAM, 99\%), N, N$^{\prime}$-methylenebisacrylamide (MBA, 99.5\%), sodium dodecyl sulfate (SDS), potassium persulfate (KPS) (99.9\%), potassium chloride (KCl, $\le$99.0\%) {and} sucrose ($\le$99.5\% ) were procured from Sigma-Aldrich. Sodium chloride (NaCl, purity 99\%) was purchased from Labort Fine Chem Pvt. Ltd. All the chemicals were used as received without any further purification.  
 
 The synthesis process followed in this study was the same as reported in a previous work \cite{mcphee1993poly, varga2001effect}. 7.0 gm NIPAM, 0.7 gm MBA and 0.03 gm SDS were dissolved in 470 ml Milli-Q water (Millipore Corp.) in a three-necked round-bottom (RB) flask. The solution was then stirred at 600 RPM and purged with N$_{2}$ gas for 30 minutes to create an inert environment for subsequent polymerization. The temperature inside the RB flask was raised to 70$^{\circ}$C, after which 0.28 gm KPS dissolved in 30 ml Milli-Q water was introduced. The introduction of KPS initiated a polymerization reaction that continued for 4 hours. The suspension was then cooled down to room temperature. Four successive rounds of centrifugations at 20,000 RPM for 60 minutes each were conducted to remove unused SDS, the remaining monomers, and other impurities. After every round of centrifugation, the supernatant was discarded and replaced by fresh Milli-Q water. After the fourth round of centrifugation, the samples were completely dry. The dried particles were ground into fine powder for further use. The spherical PNIPAM particles synthesized using this process have a core-shell structure, with the inner core exhibiting high polymer density and the peripheral shell having low polymer density \cite{varga2001effect,fullbrandt2013dielectric}.

Dense aqueous PNIPAM suspensions were prepared by adding dry PNIPAM powder to Milli-Q water. To prepare PNIPAM-additive mixtures, aqueous solutions of  additive{s of specific} concentrations were first prepared by adding NaCl, KCl or sucrose in water. A fixed amount of dry PNIPAM powder was next added to the additive solutions. All the dense suspensions thus prepared were stirred very gently over a period of 48 to 72 hours to ensure homogeneity and stored in a refrigerator at a temperature of 4$^{\circ}$C for future use. 
}

\subsection{Differential scanning calorimetry (DSC):}
{
Differential scanning calorimetry (DSC) (Mettler Toledo, DSC 3) measurements were performed to estimate the VPTT{s of dense PNIPAM suspensions of concentration 15.42\% w/v, prepared} with and without additives. 
A temperature ramp of 1$^{\circ}$C/min was applied to estimate heat flows in a temperature range of 20-50$^{\circ}$C. The heat flows evaluated in DSC experiments as a function of temperature, T, are plotted in Figs. S1(a-c){ of the supplementary material. The temperatures corresponding to the endothermic peaks were designated as the VPTTs and are displayed} in Fig. S2 {of the supplementary material}. The observed decrease in VPTT values with increase in additive concentration indicates  an earlier onset of hydrophobicity.
}
\subsection{Fourier transform infrared (FTIR) spectroscopy:}
{
FTIR spectra were measured using a Shimadzu IR Tracer-100 Fourier transform infrared spectrometer. PNIPAM suspensions of concentration 15.42\% w/v{, both} with and without additives{,} were loaded in a sample cell sandwiched between two zinc sulfide (ZnS) plates separated by a Teflon spacer of thickness 1 mm. The temperature of the sample cell was controlled using a water circulation unit (Julabo 300F). Background spectra containing either pure water or additive solutions (without PNIPAM particles) were acquired and subtracted from the spectra obtained for the PNIPAM suspensions by following the protocol given in a previous work~\cite{paz2004interaction}. Baseline correction, which involved flattening the baseline around the peaks of interest, was performed using the `Peak and Baseline' tool in Origin (version: 2023). 

The FTIR spectra of the suspensions are shown in Figs. S3-S6 of the supplementary material. Two prominent peaks, the amide II and amide I bands, were observed at ${\sim}$1560 cm$^{-1}$ and ${\sim}$1625 cm$^{-1}$, respectively. The amide II band results from N–H bending and C–N stretching vibrations in PNIPAM particles~\cite{maeda2000change}, while the amide I peak results from C=O stretching and H-bonding interactions between water and the amide moieties in PNIPAM particles ~\cite{maeda2000change}. When the temperature is raised above the VPTT, a weak peak observed at ${\sim}$ 1652 cm$^{-1}$, which represents intra or inter-molecular hydrogen bonding among the amide moieties of the PNIPAM particles~\cite{cheng2006lls}, becomes increasingly prominent. An increase in hydrophobicity due to enhanced amide-amide bonding is expected to increase the area under the weak peak at ${\sim}$ 1652 cm$^{-1}$~\cite{maeda2000change,cheng2006lls}. Simultaneously, a decrease in hydrogen bonding between the amide moiety of PNIPAM particles and water reduces the area under the ${\sim}$ 1625 cm$^{-1}$ peak~\cite{maeda2000change,cheng2006lls}.
Following the procedure adopted in previous reports~\cite{sagle2009investigating, cheng2006lls}, we deconvoluted the amide I peak obtained in FTIR data {acquired from} PNIPAM suspensions with and without additives. The areas under ${\sim}$ 1625 cm$^{-1}$ and ${\sim}$ 1652 cm$^{-1}$ peaks were estimated using the `Multiple peak fit' option of the `Peak and Baseline' tool in Origin (version: 2023). A parameter $f_{A}$, defined as the ratio of the area under the 1652 cm$^{-1}$ peak {and} the total area under the 1652 cm$^{-1}$ and 1625 cm$^{-1}$ peaks combined, is used to quantify hydrophobicity following a previous study ~\cite{cheng2006lls}. {In our analyses, plotted in Fig.~\ref{1}(a),} a higher $f_A$ indicates greater hydrophobicity.
}

\subsection{Zeta potential:}
{
Zeta potentials were measured using an electro-acoustic device (DT-100) procured from Dispersion Technology Inc.~\cite{dukhin2010characterization}. Aqueous suspensions of PNIPAM particles at a fixed particle concentration of 1\% w/v with varying additive concentrations were studied. The samples were kept inside an oil bath to maintain a fixed temperature during the experiments. The zeta potential probe comprises a transducer generating ultrasound waves at a frequency of 3 MHz. The charges present in the electrical double layer of the PNIPAM particles execute oscillations in response to the ultrasound waves, thereby producing a colloidal vibration current (CVI). The probe detects the amplitude and phase of this CVI and computes zeta potential using the DT-1200 software. Further details about the instrument can be found elsewhere~\cite{dukhin2010characterization}.  Zeta potential data acquired for PNIPAM suspensions is plotted in Fig.~\ref{1}(b).
}

\subsection{Dynamic light scattering (DLS):}
{
The mean hydrodynamic diameters, $<d_{h}>$, of non-interacting PNIPAM particles in dilute aqueous suspensions were evaluated as a function of temperature \textit{\textit{via}} dynamic light scattering experiments using a Brookhaven Instruments Corporation (BIC) BI-200SM spectrometer. The temperature of the sample was controlled between 18${^\circ}$C and 50${^\circ}$C by a water-circulation temperature controller unit (Polyscience Digital). Intensity autocorrelation functions, acquired by an autocorrelator (BI-9000AT) over a range of delay times spanning from 1 $\mu$s to 500 $\mu$s for an experimental duration of 4 minutes, were acquired and analyzed to compute $<d_{h}>$. The procedure adopted to estimate $<d_{h}>$ is given in section ST3  of the supplementary material.
Further details about the setup are given elsewhere~\cite{saha2014investigation, behera2017effects}. The mean hydrodynamic diameters, $<d_{h}>$, were further used to calculate temperature-dependent changes in the effective volume fraction,  $\phi_{eff}$, of the compressible and deformable PNIPAM particles in dense suspensions. The detailed procedure used to estimate $\phi_{eff}$ is given in section ST4 of the supplementary material. Fig. S9 of the supplementary material shows the variation of $\phi_{eff}$ as a function of temperature in the presence of various additives.  {We note a reduction in the effective volume fractions of PNIPAM suspensions upon the inclusion of additives in the aqueous medium.}
}  

\subsection{Rheology:}
{
Rheological measurements were performed using a stress-controlled MCR 702 rheometer. A cone-plate geometry with a measuring gap of 0.048 mm and a Peltier unit capable of controlling temperatures between 0$^{\circ}$C and 180$^{\circ}$C were used. Silicone oil of viscosity 5 cSt was used as a solvent trap oil to prevent solvent loss from the samples during the experiments. The viscoelastic properties of dense suspensions of PNIPAM particles of concentration 15.42\% w/v, with and without additives, were investigated by performing oscillatory rheological experiments. Temperature sweep experiments were conducted by increasing the temperature from 15$^{\circ}$C to 60$^{\circ}$C at a rate of 1$^{\circ}$C/min while maintaining the peak-to-peak strain amplitude at 0.5\% and the applied angular frequency at 0.5 rad/sec. For frequency sweep experiments, the peak-to-peak strain amplitude was kept constant at 0.5\% while logarithmically ramping up the applied angular frequency. For amplitude sweep experiments, the applied angular frequency was kept constant at 0.5 rad/sec, while the applied strain amplitude was increased logarithmically. Frequency and amplitude sweep experiments were performed at discrete temperatures below (18$^{\circ}$C), near (27$^{\circ}$C and 34$^{\circ}$C) and above (45$^{\circ}$C) the VPTT.

The structural relaxation timescales of soft glassy systems are very slow and usually lie outside the experimental time window accessible in a regular frequency sweep test~\cite{romeo2010temperature,minami2016rheological}. To quantify the slow glassy dynamics of the samples below the VPTT, strain rate frequency sweep (SRFS) experiments~\cite{wyss2007strain, mohan2008phase} were performed. In contrast to regular frequency sweep measurements where the oscillatory strain amplitude, ${\gamma_{0}}$, is kept constant, the strain rate amplitude, $\dot{\gamma_{0}}$, is kept constant in SRFS experiments. 
It was shown that under an externally imposed strain, a system's structural relaxation timescale satisfies the empirical relation 1/$\tau$($\dot{\gamma_{0}}$) = 1/$\tau_{0}$ + K$\dot{\gamma_{0}}^\nu$, where $\tau_{0}$ is the relaxation timescale in the absence of external strain, $\tau$($\dot{\gamma_{0}})$ is the structural relaxation timescale measured upon the application of a strain rate amplitude $\dot{\gamma_{0}}$, while K and $\nu$ are positive constants~\cite{wyss2007strain}. By applying sufficiently large $\dot{\gamma_{0}}$ values, relaxation processes can be {sped up to lie} within the experimentally accessible frequency range. As a result, even ultra-slow relaxation processes of soft glassy materials can be reliably determined in SRFS experiments.    
}

\subsection{Cryogenic field emission scanning electron microscopy (cryo-FESEM):}
{
Cryo-FESEM was carried out using a field emission scanning
electron microscope (Ultra Plus FESEM-4098; Carl Zeiss) having an electron beam strength of 5 KeV. The samples were first equilibrated to a pre-determined temperature and subsequently loaded in a rivet fixed with a cryostub. The samples were then vitrified in liquid nitrogen slush at -190$^{\circ}$C 
and transferred to a vacuum chamber maintained at -190$^{\circ}$C (PP3000T Quorum Technology). The samples were next sublimated at -90$^{\circ}$C for 5 minutes and cryo-fractured using an in-built stainless steel knife. The fractured samples were further sublimated for 12 minutes at -90$^{\circ}$C to remove ice from the sample structures. A gold coating of thickness around 5 nm was then applied to the sample surfaces to ensure good image contrast. Finally, the samples were transferred to the cryo-chamber, maintained at a temperature of -190$^{\circ}$C, for imaging. Back-scattered secondary electrons were used to reconstruct the surface images of the samples.   
}

\section{\label{r&d}Results and Discussion}
\subsection{Modifying hydrophobicity and electrostatic interactions in aqueous PNIPAM suspensions by incorporating additives:}

\begin{figure}[th]
     \centering
     \includegraphics[width=1\linewidth]{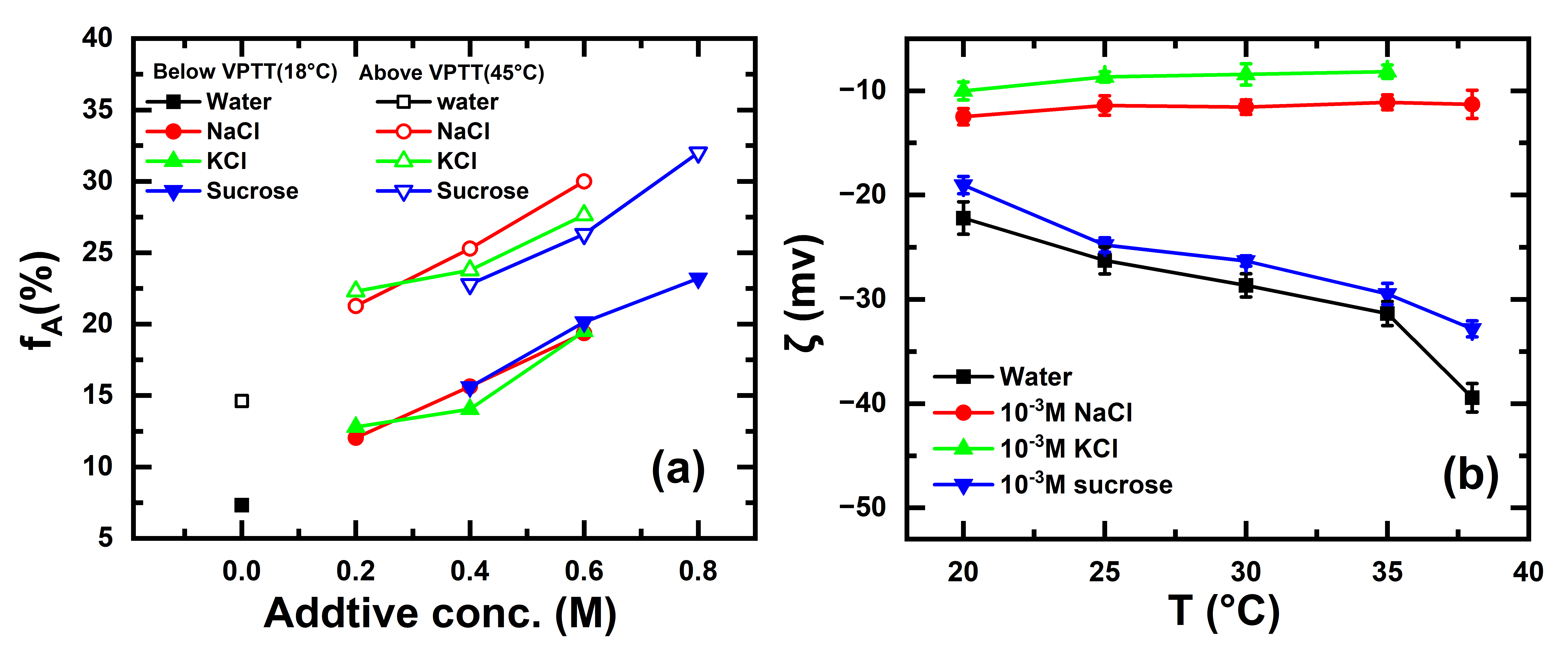}
     \caption{(a) The relative strength of the amide-amide bonding, $f_A$, used here to parametrize particle hydrophobicity, is plotted as a function of additive concentration { for NaCl (red), KCl (green) and sucrose (blue) } both below (18$^{\circ}$C, bold symbol{s}) and above (45$^{\circ}$C, hollow symbol{s}) the volume phase transition temperature (VPTT). (b) Temperature-dependent zeta potential of 1\% w/v PNIPAM particles suspended
      in water and in different additive solutions.
     }
     \label{1}
 \end{figure}

The alteration in particle hydrophobicity due to the incorporation of additives in aqueous PNIPAM suspension was quantified by performing FTIR measurements as discussed in section II.3. Fig.~\ref{1}(a) shows that the measure of hydrophobicity, {$f_{A}$,} increases with both temperature and additive concentration. 

{Previous e}xperiments and MD simulations have reported that the introduction of osmolytes like sucrose in an aqueous medium increases the hydrophobicity of PNIPAM particles~\cite{narang2017unexplored}.  {Sucrose, being highly hydrophilic, displaces water molecules from the hydration shell of the particles.} Dehydration in the presence of sucrose facilitates intramolecular hydrogen bonding in the PNIPAM particles, leading to a coil to globule transition~\cite{narang2017unexplored}.
Additionally, a minor contribution to hydrophobicity also comes from direct interactions between the amide moieties of the particle and sucrose, which reduces the number of sites available for hydrogen bonding between water molecules and the PNIPAM particles~\cite{narang2017unexplored}. This results in an increase in particle hydrophobicity with an increase in sucrose concentration as seen in Fig.~\ref{1}(a).

In contrast, the presence of ionic salts like NaCl and KCl in the aqueous medium enhances hydrophobicity mainly by increasing the surface tension at the particle-water interface~\cite{zhang2005specific, zhang2010chemistry}. Water molecules form ordered structures around the hydrophobic polymer backbone and the isopropyl groups of the PNIPAM particles, a phenomenon known as hydrophobic hydration. Cl$^{-1}$ anions raise the surface tension at the interface formed by the water molecules and the hydrophobic groups in the PNIPAM particles ~\cite{zhang2005specific, zhang2010chemistry}. This inhibits hydrophobic hydration, leading to a significant enhancement in particle hydrophobicity. Previous studies have reported that surface tension at the aqueous-polymer interfaces increases with an increase in salt concentration~\cite{pegram2006partitioning}. This is consistent with our FTIR data presented in Fig.~\ref{1}(a), {which clearly shows} that hydrophobicity increases with an increase in salt concentration. Hydrophobicity also increases with temperature, regardless of the polarity of the additive in the suspension medium.

The impact of additives on inter-particle electrostatic interactions was also analyzed through zeta potential measurements, as described in section II.4. Fig.~\ref{1}(b) shows temperature-dependent variations of the surface zeta potentials, $\zeta$, of PNIPAM particles in aqueous suspensions both in the presence and absence of dissociating and non-dissociating additives. 
PNIPAM particles acquire surface charges due to the use of a surfactant (SDS) and initiator (KPS) during particle synthesis. As the temperature rises and the particles shrink, the surface charge density increases~\cite{town2019understanding, misra2024effect}, leading to higher negative zeta potential values in PNIPAM samples prepared in pure water and non-dissociating sucrose solutions. 
Previous reports have shown that a minimum zeta potential of $\pm$30mV is required for electrostatic stabilization of colloidal particles in suspension~\cite{hunter2013zeta}.
The high negative zeta values of PNIPAM suspensions prepared in pure water and sucrose solution above the VPTT therefore indicate good electrostatic stability. In the presence of NaCl and KCl, however, the Debye screening layers surrounding colloidal PNIPAM particles shrink due to an increase in the number of dissociated ions in the suspension medium.
Addition of dissociating salts, therefore, results in the observed weakening of $\zeta$, which significantly accelerates the flocculation of PNIPAM particles~\cite{rasmusson2004flocculation,daly2000study} \textit{via} electrostatic attraction even in dilute suspensions.

\subsection{Influence of additives on the temperature-dependent viscoelastic properties of aqueous PNIPAM suspensions:}

\begin{figure*}
     \centering
     \includegraphics[width=0.4\linewidth]{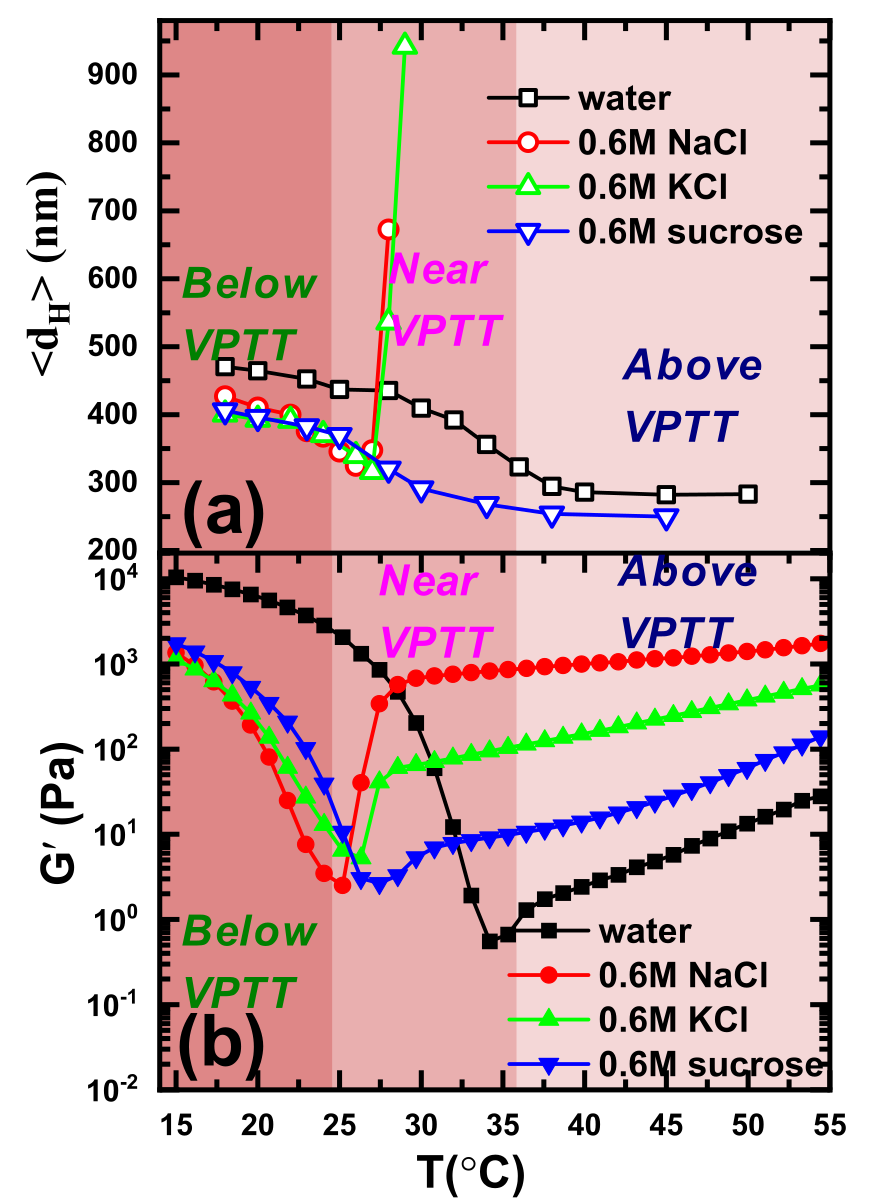}
     \caption{Temperature-dependent {(a)} mean hydrodynamic diameters, $<d_{H}>$, and {(b)} storage moduli, {G$^{\prime}$}, of PNIPAM particles suspended in {water and in} various additive solutions.}
     \label{2}
 \end{figure*}

Fig.~\ref {2}(a) displays representative data of the mean hydrodynamic diameter{s}, $<d_{h}>$, {as a function of temperature, for} PNIPAM particles suspended in pure water and also in 0.6 M NaCl, KCl and sucrose. The data for temperature-dependent $<d_{h}>$ values for all the other samples used in this study are plotted in Figs. S10(a-c) of the supplementary material. 
Figs.~\ref{2}{(a)} and S10(a-c) show a systematic reduction in $<d_{h}>$ {below the VPTT} with increasing additive concentration in the suspension medium. For suspensions prepared in an ionic medium with salts like NaCl and KCl, {we note} an abrupt rise in $<d_{h}>$ near the VPTT, as displayed in Figs.~\ref{2}{(a)} and S10(a,b). 

Enhanced hydrophobicity in the presence of both dissociating and non-dissociating additives, as seen from our FTIR data in Fig.~\ref{1}{(a)}, leads to {particle} shrinkage. This, in turn, results in a reduction of effective volume fraction, $\phi_{eff}$,  of the suspensions, as depicted in Fig. S9.  Meanwhile, the abrupt rise in $<d_{h}>$ near the VPTT {for samples with salts, displayed in {Fig.~\ref{2}{(a),}} can be explained in terms of particle flocculation due to {the combined effects of} electrostatic attraction, as {suggested} by the weak $\zeta$ values in the presence of salts,  and high particle hydrophobicities {(Fig.~\ref{1})}}. In contrast, the higher {negative} $\zeta$ values {of samples prepared in} pure water and sucrose near the VPTT {indicate} particle stability against electrostatic aggregation, and thus we do not see any sudden jump in $<d_{h}>$ in {these experiments}. {Although} $\zeta$ values are low below the VPTT, our DLS data do not show any signature{s} of flocculation in this temperature regime. {This arises due to the negligible} mismatch {between} the Hamaker constant{s of} the swollen PNIPAM particles and the aqueous medium~\cite{rasmusson2004flocculation}.

Fig.~\ref{2}(b) shows a representative plot of the elastic moduli, G$^{\prime}$, as a function of temperature, for suspensions of PNIPAM particles{ prepared} with and without additives {in the aqueous medium}. The temperature-dependent {G$^{\prime}$}  values for all the other samples used in this study are plotted in Figs.~S10(d-f), wh{ile} the corresponding viscous moduli, {G$^{\prime \prime}$}, values are shown in Figs.~S11(a-c) of the supplementary material. We note from Fig.~\ref{2}(b) that {G$^{\prime}$} values {decrease} {with increasing concentrations of} dissociating and non-dissociating additives {below the VPTT}. 
{Hydrophobic shrinkage of PNIPAM particles  due to the inclusion of additives and the resultant decrease in suspension volume fraction, as} shown in Figs.~\ref {1}(a),~\ref {2}(a) and S9, result in the observed loss in rigidity below the VPTT. 

Near the VPTT, we observe that {G$^{\prime}$} values increase rapidly in the {samples with} dissociating salts at temperatures that correlate well with the abrupt jumps in $<d_{H}>$. 
Clearly, the observed sharp increase in suspension rigidity is a consequence of the aggregation of PNIPAM particles into clusters due to enhanced electrostatic and hydrophobic inter-particle attractions under these conditions. {While the samples prepared in water or sucrose solutions may presumably form weak interconnected particle clusters due to increased particle hydrophobicity near the VPTT, we note that their} low G$^{\prime}$ values verify the absence of flocculation driven by electrostatic attraction.  

Above the VPTT, aqueous suspensions of PNIPAM microgels are known to form attractive gels~\cite{romeo2010temperature, minami2020criteria}. Attractive inter-particle interactions, originating from growing particle hydrophobicity, are believed to drive the gelation process~\cite{misra2024effect,minami2016rheological}. 
Contrary to our observations below the VPTT, we find that suspension rigidities are higher {above the VPTT} in the presence of additives. 
We further note from Figs.~\ref{2}(b) and S10(d-f) that {G$^{\prime}$} values of suspensions containing sucrose are higher than those prepared in pure water, but lower than those with {dissociating} additives in the suspension medium. The build-up of moduli values is also much more gradual in {samples with} sucrose. This {verifies our earlier intuition} 
that while hydrophobicity-induced particle shrinkage governs bulk rheological characteristics of aqueous PNIPAM suspensions in the presence of additives below the VPTT, both hydrophobicity and electrostatics {influence} suspension properties near and above the VPTT.

\subsection{Influence of additives on particle dynamics and {suspension} rheolog{y} below the VPTT:} 
 
\begin{figure*}
     \centering
     \includegraphics[width=0.8\linewidth]{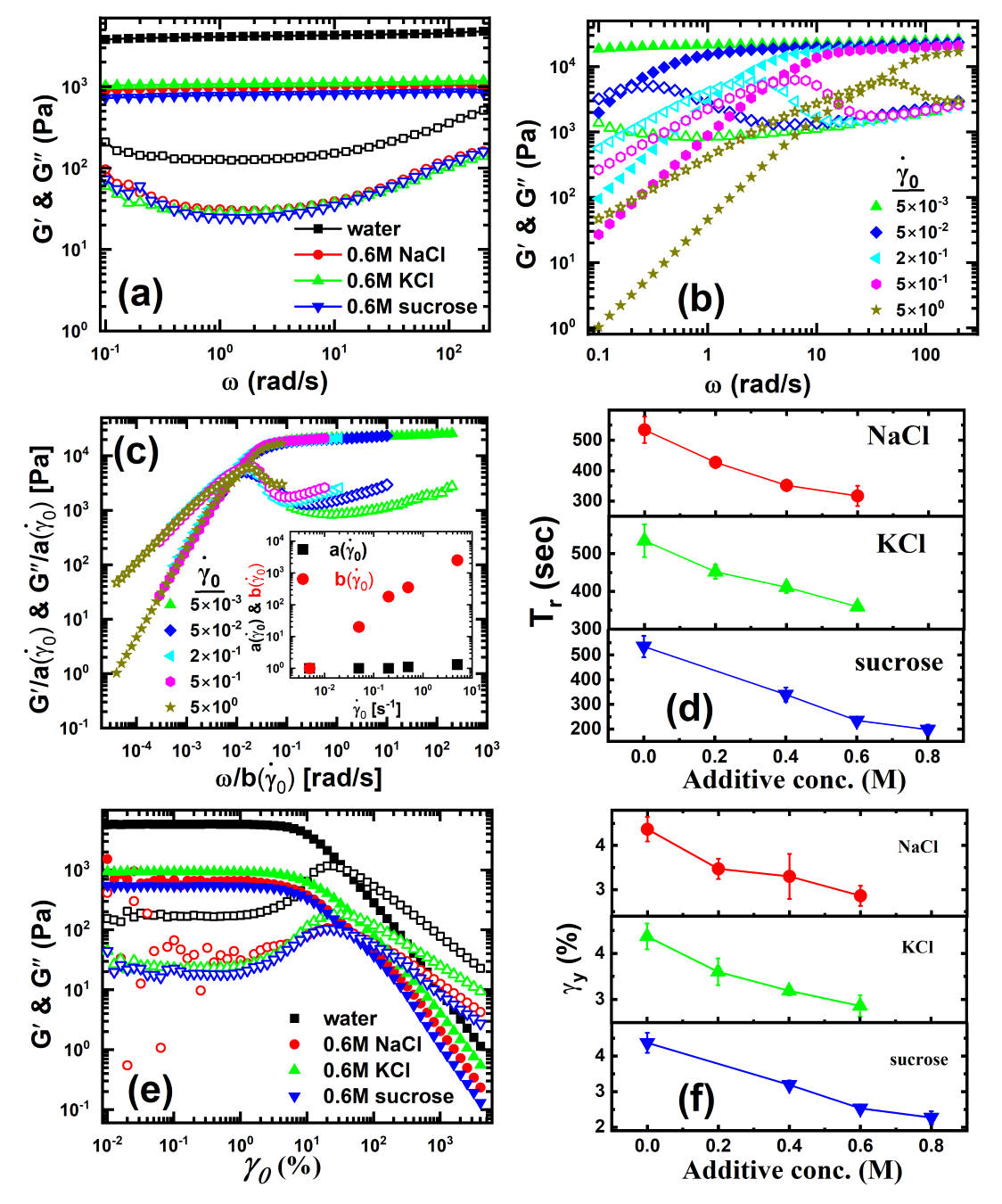}
     \caption{(a) Frequency-dependent elastic moduli {G$^{\prime}$}  (solid symbols) and viscous moduli {G$^{\prime \prime}$} (hollow symbols), measured at 18$^{\circ}$C (below the VPTT), of dense suspensions of PNIPAM particles prepared in different additive solutions. (b) {G$^{\prime}$}  (solid symbols) and {G$^{\prime \prime}$} (hollow symbols) of PNIPAM particles suspended in pure water as a function of $\omega$ for five different $\dot{\gamma_{0}}$ at 18$^{\circ}$C. (c) Data shown in (b) is scaled and collapsed on a single master curve. The inset shows the scaling parameters. (d) Structural relaxation timescale{s} of dense suspensions of PNIPAM particles as a function of additive concentration at 18$^{\circ}$C. (e) Strain amplitude dependent elastic moduli, {G$^{\prime}$},  (solid symbols) and viscous moduli, $G^{\prime\prime}$, (hollow symbols) of dense suspensions of PNIPAM particles at 18$^{\circ}$C. (f) Yield strains of dense suspensions of PNIPAM particles as a function of additive concentration estimated from (e).
     }
     \label{3}
 \end{figure*}

Fig.~\ref{3}(a) shows a representative plot of frequency sweep oscillatory rheology data of PNIPAM particles suspended in pure water and different additive solutions below the VPTT (18$^{\circ}$C). 
We observe that {G$^{\prime}$}  exhibits very weak or no angular frequency ($\omega$) dependence, while {G$^{\prime \prime}$} shows a minimum at a characteristic frequency. Furthermore, {G$^{\prime}$}  > {G$^{\prime \prime}$} over the frequency range explored. These are typical features of soft glasses~\cite{mason1995elasticity,pham2008yielding,koumakis2012direct}. 
Additional frequency sweep data below the VPTT are shown in Figs. S12(a-c) of the supplementary material. It can be clearly seen from Figs.~\ref{3}(a) and S12(a-c) that the moduli values decrease significantly with the introduction of additives, suggesting loss of mechanical rigidity, as previously seen in our temperature sweep data below the VPTT.
Additionally, we note that {G$^{\prime}$}  and {G$^{\prime \prime}$} do not cross over in the frequency range explored. This indicates that the structural relaxation timescales, $T_r$, are extremely slow and lie outside the experimental frequency window. To estimate $T_r$, we conducted strain rate frequency sweep (SRFS) experiments{~\cite{wyss2007strain, mohan2008phase},} wherein pre-determined strain rate amplitudes, $\dot{\gamma_{0}}$,  were applied to drive the structural relaxation process to experimentally accessible frequencies, as discussed in section II.6. Previously, SRFS has been successfully used to access the slow structural relaxation processes of soft microgel systems comprising cross-linked PNIPAM and polyacrylic acid~\cite{wyss2007strain} and for micelle forming triblock copolymers~\cite{mohan2008phase}.   
Fig.~\ref{3}(b) shows the variations of {G$^{\prime}$}  and {G$^{\prime \prime}$} with $\omega$ for PNIPAM particles suspended in pure water at five different $\dot{\gamma_{0}}$. The corresponding data for samples containing additives are shown in Figs. S13(a-c)-S15(a-c) of the supplementary material. 

We note from Figs.~\ref{3}(b) and S13(a-c)-S15(a-c) that the {crossover} frequencies of {G$^{\prime}$}  and {G$^{\prime \prime}$} shift towards higher values of $\omega$ with increase in $\dot{\gamma_{0}}$. By using the relation $G_{scaled}^{*}(\omega)$ = $\frac{G^{*}(\omega/b(\dot{\gamma_{0}}))}{a(\dot{\gamma_{0}})}$~\cite{wyss2007strain}, where $G^{*}$ is either {G$^{\prime}$}  or {G$^{\prime \prime}$}, ${a(\dot{\gamma_{0}})}$ and ${b(\dot{\gamma_{0}})}$ are the scaling factors for the moduli and the frequencies, respectively, the data corresponding to different $\dot{\gamma_{0}}$ were appropriately scaled to collapse on a single master curve as shown in Figs.~\ref{3}(c) and S13(d-f)-S15(d-f).
Fig.~\ref{3}(d) shows the characteristic structural relaxation timescale, estimated as $T_r=2\pi/\omega{_o}$, where $\omega{_o}$ corresponds to the crossover between the scaled {G$^{\prime}$}  and {G$^{\prime \prime}$} data. {When dissociating and non-dissociating additive concentrations were increased in the suspension medium,} we observe that $T_r$ decreased systematically, indicating faster particle dynamics. 
    
Fig.~\ref{3}(e) shows plots of amplitude sweep experiments performed at 18$^{\circ}$C for {the same} PNIPAM {suspensions as in Fig.~\ref{3}(a)}. Amplitude sweep data acquired below the VPTT for all the samples are displayed in Figs. S16(a-c) of the supplementary material.
It can be seen from Figs.~\ref{3}(e) and S16(a-c) that {for small $\gamma_0$,} {G$^{\prime}$}  and {G$^{\prime \prime}$} are independent of the applied $\gamma_0$ {and} {G$^{\prime}$}  > {G$^{\prime \prime}$}. With increase in $\gamma_0$, {G$^{\prime}$}  starts to decrease monotonically, wh{ile} {G$^{\prime \prime}$} shows a peak before decreasing. These features are {consistent with our earlier observations in Fig.~\ref{3}(a) and} are reminiscent of soft glassy rheology~\cite{pham2008yielding, mason1995elasticity}. {As expected}, the mechanical moduli estimated from amplitude sweep experiments also decrease systematically with increase in additive concentration. Interestingly, the {G$^{\prime \prime}$} peak heights decrease with additive content as seen from Figs.~\ref{3}(e) and S16(a-c), indicating a reduction in inter-particle repulsion with increasing particle hydrophobicity\cite{pham2008yielding}.
We estimated the yield strains, $\gamma_y$, defined as the strain values that {signal} the onset of the yielding process, {by} following a procedure described in detail in section ST8 of the supplementary material. Fig.~\ref{3}(f) {displays that} $\gamma_y$ {decreases} with increase in concentrations {of both dissociating and non-dissociating additives}. 

We further note from Fig.~S18 of the supplementary material that {a direct correlation exists between} $T_r$ {and} $\gamma_y$, suggesting that {a common parameter governs} the behavior{s} exhibited by both parameters. Hydrophobicity-induced particle shrinkage in the presence of additives {below the VPTT} creates additional free volume in the suspension medium. This {is apparent from Fig. S9, with the observed decrease in suspension effective volume fraction with increasing additive content indicating the creation of} free volume{. The creation of excess free volume} leads to accelerated diffusion of particles out of the{ir} cages {and} yielding at relatively smaller deformations, resulting in the {simultaneous} lowering of $T_r$ and $\gamma_y$. Additionally, the reduction in {moduli values} and yield strain with an increase in additive content also suggests that enhanced hydrophobicity renders the suspensions {mechanically weaker} and more brittle\cite{carrier2009nonlinear}. 
Previously, we ha{d} seen from Fig.~\ref{1}(a) {and Figs.~\ref{3}(d,f)} that the concentration-dependence{s} of the hydrophobicity parameter, $f_{A}$, {the structural relaxation time $T_r$ and the yield strain $\gamma_{y}$ are} very similar {for suspensions prepared with} both dissociating and non-dissociating additives. 
{These observations, along with the observed hydrophobic shrinkage of the particles below the VPTT, as seen in Fig.~\ref{2}(a), allow us to} conclude that additive-induced hydrophobicity governs the rheological response below the VPTT, regardless of whether the additives are {dissociating or non-dissociating}. Finally, we also note that {the influence of} inter-particle electrostatic interactions {on} suspension rheology {is negligible} below the VPTT .

\subsection{Influence of additives on particle dynamics and {suspension} rheolog{y} near the VPTT:}

\begin{figure*}
     \centering
     \includegraphics[width=0.6\linewidth]{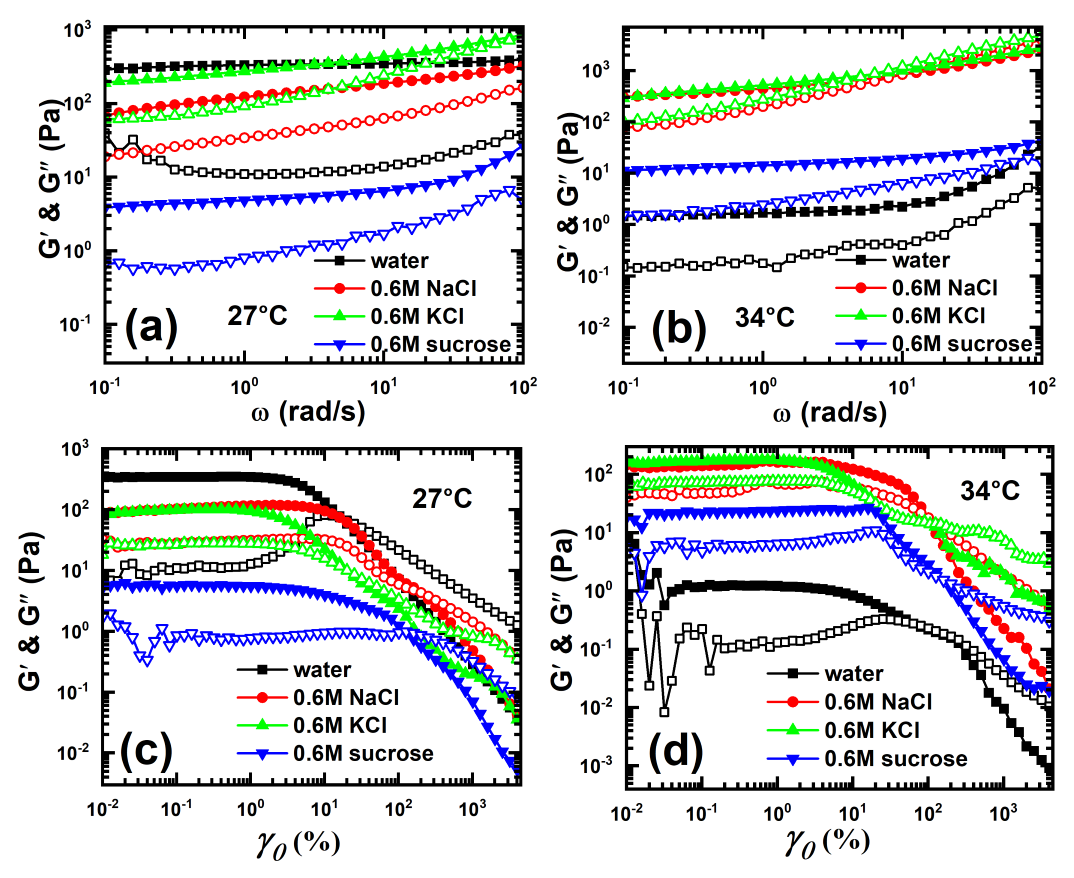}
     \caption{Frequency-dependent elastic moduli {G$^{\prime}$}  (solid symbols) and viscous moduli {G$^{\prime \prime}$} (hollow symbols), of dense suspensions of PNIPAM particles{, prepared in pure water and in additive solutions,} at (a) 27$^{\circ}$C and (b) 34$^{\circ}$C. Strain amplitude-dependent elastic moduli {G$^{\prime}$}  (solid symbols) and viscous moduli {G$^{\prime \prime}$} (hollow symbols) of {the same} suspensions at (c) 27$^{\circ}$C and (d) 34$^{\circ}$C. 
     }
     \label{4}
 \end{figure*}

Frequency and amplitude sweep experiments with PNIPAM suspensions, prepared in pure water or in solutions containing dissociating and non-dissociating additives, were next performed at temperatures near the VPTT. Since the VPTT decreases in the presence of additives, {two temperatures that lie close to the VPTTs of these suspensions,} 27$^{\circ}$C and 34$^{\circ}$C, were selected. 
Fig.~\ref{4}(a) displays the results of frequency sweep experiments conducted at 27$^{\circ}$C. We see that the PNIPAM suspension prepared in pure water continues to exhibit features typical of soft glassy rheology at 27$^{\circ}$C, but with much lower moduli value{s} compared to those at 18$^{\circ}$C (plotted in Fig.~\ref{3}(a)) {due to hydrophobic shrinkage of the constituent PNIPAM particles}. This behavior is not surprising as 27$^{\circ}$C lies below the VPTT of pure aqueous PNIPAM suspension{s} ($\sim$33$^{\circ}$C) as previously shown in Fig.~S2. 
In the presence of dissociating additives, we note from Fig.~\ref {4}(a) that both {G$^{\prime}$}  and {G$^{\prime \prime}$} show a{pproximately} power-law dependence{s} on the applied frequency, $\omega$, {signaling the onset of} colloidal gel {formation}~\cite{prasad2003rideal}. We note that 27$^{\circ}$C lies slightly above the VPTTs ($\sim$26$^{\circ}$C) of PNIPAM suspensions prepared in additive solutions.

Fig.~\ref{4}(b) shows that the rheology of all the additive-containing samples {at 34$^{\circ}$C} are characterized by {higher moduli values than those reported in Fig.~\ref{4}(a)}, reflecting {the rapid formation of gels just above the VPTT}~\cite{prasad2003rideal,minami2018viscoelasticity}. While the samples prepared in dissociating salt solutions display {the} high{est} moduli values and {the} strong{est} $\omega$-dependences, those prepared in pure water exhibit the lowest moduli values {and the weakest $\omega$-dependence}. {Since hydrophobicity is the weakest f}or PNIPAM particles {suspended} in pure water as seen in Fig.~\ref{1}(a), these samples display the lowest moduli values. Additionally, stronger electrostatic stabilization in the case of pure water and sucrose, as indicated by their higher {negative} $\zeta$ values shown in Fig.~\ref{1}(b), inhibit gelation near the VPTT. In contrast, the presence of dissociating additives enhances particle flocculation due to electrostatic attraction as the {temperature} is {raised across the VPTT}. Thus, the high {G$^{\prime}$} and {G$^{\prime \prime}$} values in these samples verify our previous claim that gelation {near the VPTT} results from a combined effect of electrostatic and hydrophobic attraction for suspension prepared {in solutions containing} dissociating additives.  

Figs.~\ref{4}(c) and ~\ref{4}(d) display the amplitude sweep data of the same samples at 27$^{\circ}$C and 34$^{\circ}$C, and are consistent with the results of the frequency sweep {experiments} discussed above. We again note that the mechanical moduli values are considerably higher in the presence of dissociating additives, {verifying that} the combined effect of hydrophobic and electrostatic attractions {drive {gel formation}} near the VPTT. 
{Therefore, o}ur findings in this section suggest that unlike below the VPTT where electrostatics has a negligible impact, both hydrophobicity and electrostatics can determine the rheological features of microgel suspensions at temperatures near the VPTT. 

\subsection{Influence of additives on particle dynamics and {suspension} rheolog{y} above the VPTT:}

\begin{figure*}
     \centering
     \includegraphics[width=0.6\linewidth]{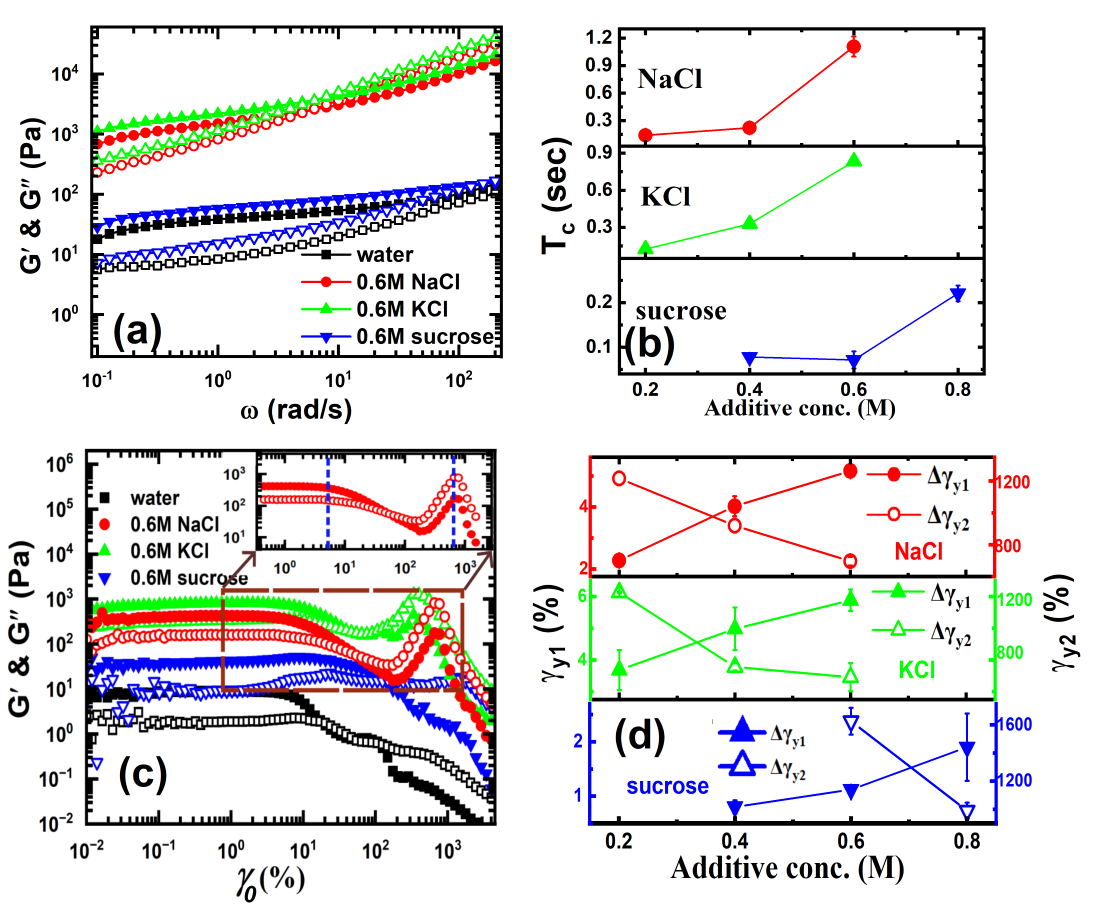}
     \caption{(a) Frequency-dependent elastic moduli, {G$^{\prime}$}  (solid symbols) and viscous moduli, {G$^{\prime \prime}$} (hollow symbols), of dense suspensions of PNIPAM particles {{prepared} with and without additives} at 45$^{\circ}$C (above the VPTT). (b) Crossover timescales of {these} suspensions as a function of additive concentration. (c) Strain amplitude-dependent elastic moduli, {G$^{\prime}$}  (solid symbols), and viscous moduli, {G$^{\prime \prime}$} (hollow symbols), of {the same samples} at 45$^{\circ}$C. The inset shows a zoomed-in {view} of the two-step yielding of {a} PNIPAM suspension containing 0.6M NaCl. (d) Yield strains, $\gamma_{y1}$ and $\gamma_{y2}$, of {the samples} at 45$^{\circ}$C as a function of additive concentration.
     }
     \label{5}
 \end{figure*}

Fig.~\ref{5}(a) displays frequency sweep experiments for aqueous PNIPAM suspensions prepared in pure water and 0.6M additive solutions at 45$^{\circ}$C, a temperature that lies well about the VPTT. The measured moduli values are significantly higher than those near the VPTT, and their strong frequency dependences are reminiscent of gel-like rheology~\cite{prasad2003rideal,minami2018viscoelasticity}. We also note that the moduli of the suspensions prepared with dissociating
additives are {almost two orders} of magnitude higher than those containing non-dissociating sucrose.
As discussed earlier, the higher moduli observed in the presence of dissociating salts arise from the combined influence of electrostatic and hydrophobic interactions. Dissociating additives therefore promote the formation of stronger and more rigid gels. For suspensions prepared in pure water and sucrose solution, the attractive inter-particle interaction comes from hydrophobicity alone, and the absence of electrostatic attraction leads to more modest changes in the moduli with applied frequency. 

Additional frequency sweep data above the VPTT are shown in Figs. S19(a-c) of the supplementary material. { We observe from Figs.~\ref{5}(a) and S19(a-c) that {G$^{\prime}$} and {G$^{\prime \prime}$} cross over at a {characteristic} frequency, $\omega_{c}$. The low frequency {rheological} response ($\omega$ < $\omega_{c}$, where {G$^{\prime}$} > {G$^{\prime \prime}$}) is {expected to be} dominated by rigid gel networks\cite{prasad2003rideal,gardel2004scaling}. For the frequency regime $\omega$ > $\omega_{c}$, we believe that the response is mainly governed by fluctuations in the gel strands\cite{gardel2004scaling}. Fig.~\ref{5}(b) shows that the characteristic relaxation timescales, $T_c$, estimated as $T_c= 2\pi/\omega_{c}$, increase monotonically with an increase in additive concentration.}
We further note that $T_c$ is slower for samples made in solutions of dissociating salts. This verifies that the additional contribution of electrostatic attraction significantly alters particle mobility in the gel phase above the VPTT.   

Fig.~\ref{5}(c) displays representative plot{s} of amplitude sweep experiments conducted above the VPTT (45$^{\circ}$C). The raw data for all the samples are shown in Figs.~S20(a-c) of the supplementary material. In contrast to single-step yielding seen below the VPTT, we observe a two-step yielding behavior, a typical feature of attractive systems~\cite{pham2008yielding,laurati2011nonlinear,koumakis2011two,ahuja2020two}, in the presence of additives.  We further observe from Figs.~\ref{5}(c) and S20(a-c) that the mechanical moduli {are the highest} for suspensions containing {dissociating} salts, {confirming} the strong contribution of electrostatic attraction in enhancing the rigidities of the gel networks that form {above the VPTT}. 

Fig.~\ref{5}(d) shows a plot of the first and the second yield strain{s}, $\gamma_{y1}$ and $\gamma_{y2}$ respectively, as a function of additive concentration above the VPTT, estimated by following a procedure outlined in section ST11 of the supplementary material. We observe that $\gamma_{y1}$ increases but $\gamma_{y2}$ decreases with increase in additive concentration. 
The first yield strain, $\gamma_{y1}$, is attributed to {bond-breaking events between particles within the gel network}~\cite{koumakis2011two}.
The attractive strength between particles increases with higher additive content, as confirmed by the elevated mechanical moduli values of the suspension in Figs.~\ref{5}(c) and S20(a-c).
Thus, { gel strand fluctuations should reduce} as the additive concentration rises, {necessitating} { higher strains  for bond breakage. This leads to the observed rise in $\gamma_{y1}$ with increase in additive concentration.} 
The direct correlation between $T_c$, obtained from oscillatory frequency response rheology, and $\gamma_{y1}$, {obtained from amplitude sweep rheology}, is displayed in Fig. S22 of the supplementary material and further verifies the {sensitive} dependence of suspension rheology on inter-particle interactions and particle-scale dynamics.

The second yield strain, $\gamma_{y2}$, in attractive gels emerges from the breakage of interconnected clusters into smaller fragments~\cite{misra2024effect, koumakis2011two}. 
It has been shown previously that $\gamma_{y2}$ in an attractive colloidal gel system decreases with an increase in attraction strength due to the formation of less flexible clusters~\cite{laurati2011nonlinear}. 
{Since} higher additive content{s} lead to the formation of more rigid clusters {in our systems}, {the}  smaller $\gamma_{y2}$ values required to rupture them, as shown in Fig.~\ref{5}(d){, are not surprising}. 
{T}he reduction in $\gamma_{y2}$ with {additive concentration} reveals that the suspensions {become} more brittle, with the presence of larger salt contents expected to induce system-wide fractures at smaller strains.  
{{Furthermore, reduction in $\gamma_{y2}$} is accompanied by increase in the} heights of the {G$^{\prime \prime}$} peaks, as shown in  Figs.~\ref{5}(c) and S20(a,b), indicat{ing} enhanced dissipation due to the breakup of the highly rigid clusters. These observations again clearly highlight the influence of electrostatic attraction on the rheological signatures of dense PNIPAM suspensions above the VPTT.
We conclude that the bulk rheological behavior and particle assembly above the VPTT are governed by inter-particle attraction{s} that can be tuned by controlling both hydrophobic and electrostatic interactions {\textit{via} the incorporation of suitable additives in the suspension medium}.  

\subsection{\label{se:col} Correlati{ng} rheological properties {with} suspension microstructures:}
We next attempt to correlate the rheological properties {of the suspensions} with the{ir} microscopic structure{s}. Microstructural images of aqueous PNIPAM suspensions both with and without additives were acquired using cryogenic field emission scanning electron microscopy (cryo-FESEM) as outlined in section II.7. Figs.~\ref{6}(a-d) display cryo-FESEM images of PNIPAM particles suspended in pure water and different additive solutions {whose temperatures were maintained} below the VPTT {before being cryo-frozen}. 
Fig.~\ref{6}(a) shows that PNIPAM particles assemble into dense networks with small pores when pure water is the suspension medium, as highlighted by yellow dashed {lines}.  {The self-
\begin{figure}[th]
     \centering
     \includegraphics[width=1\linewidth]{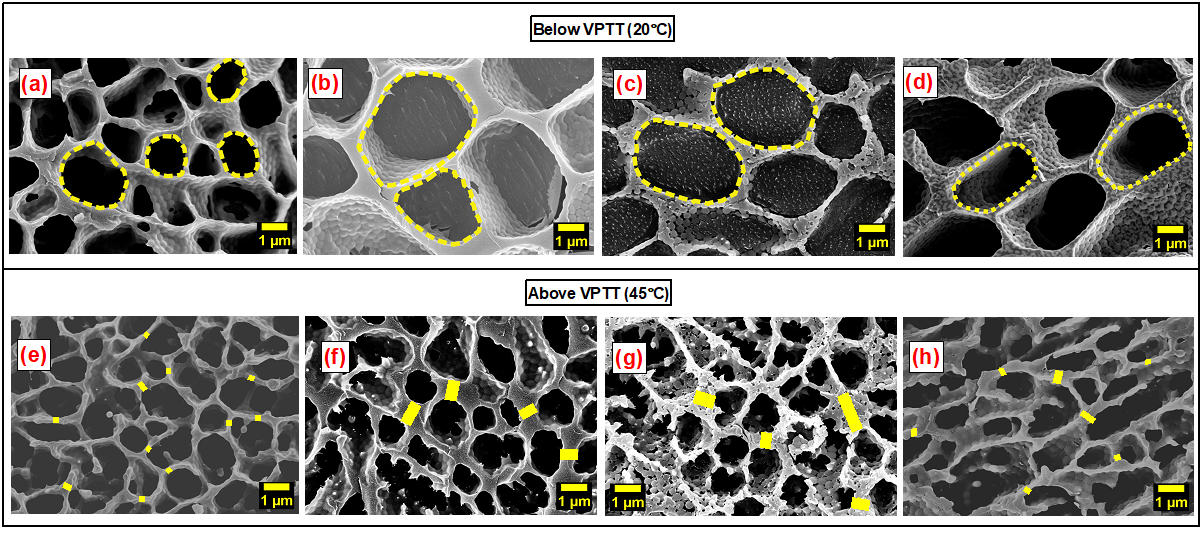}
     \caption{Cryo-FESEM images of dense suspensions of PNIPAM particles suspended in (a) pure water (b) 0.6M NaCl, (c) 0.6M KCl, (d) 0.6M sucrose, {prepared} below the VPTT (20$^{\circ}$C) {before being cryo-frozen}. Yellow dashed regions highlight some of the pores present in the suspensions. The images {of the same samples as in (a-d), prepared} above the VPTT (45$^{\circ}$C) {before being cryo-frozen, }are shown in (e), (f), (g) and (h) respectively. The yellow lines used here denote strand thickness{. The scale bars represent 1 $\mu$m}.}
     \label{6}
 \end{figure}
assembly of PNIPAM particles, synthesized with low crosslinker (MBA) content {such as in the present work}, into gel-like networks below the VPTT has been demonstrated by us in a previous} study~\cite{misra2024effect}. 
Figs.~\ref{6}(b-d) show that the pore sizes {of the inter-connected network{s} of PNIPAM particles} increase with the incorporation of additives. 
We have already shown {in Fig. S9} that effective volume fraction{s} {of the suspensions} decrease with the introduction of both dissociating and non-dissociating additives below the VPTT {due to hydrophobicity-induced particle deswelling}. 
This decrease in $\phi_{eff}$ {due to enhanced particle hydrophobicity} lead{s} to the widening of the pores in the presence of additives, as seen from Figs. ~\ref{6}(b-d).
Structures that are more porous should have less structural rigidity. Indeed, our rheology data demonstrated that PNIPAM suspensions prepared in additive solutions are characterized by lower mechanical moduli below the VPTT {when compared to suspensions prepared in pure water}.

Figs.~\ref{6}(e-h) shows cryo-FESEM images of PNIPAM particles suspended in pure water and different additive solutions. The temperatures of the samples were raised above the VPTT before they were cryo-frozen. 
We observe porous, polydisperse space-spanning {attractive gel networks}. {As shown by the yellow lines in Figs.~\ref{6}(f) and ~\ref{6}(g)}, denser networks with thicker gel strands {form} in the presence of dissociating salts, indicating enhanced gelation.
In contrast, suspensions prepared in pure water {and sucrose}, shown in Figs. ~\ref{6}(e) and ~\ref{6}(h) respectively, {form gel-like structures that are characterized} by large pores with thinner strands. These observations {highlight} the {significant} influence of {flocculation} {induced by electrostatic attractions} in {driving} {the} gelation {of PNIPAM suspensions prepared in salt solutions} above the VPTT, and are consistent with our observation of higher moduli values {under these conditions}.   

\section{\label{se:sac} Conclusions:}
The combined effects of hydrophobicity and electrostatics in determining the bulk rheological properties of dense aqueous suspensions of PNIPAM particles is investigated in this study. {D}issociating (NaCl and KCl) and non-dissociating (sucrose) additive{s} were added to dense aqueous suspensions of PNIPAM {microgel} particles to tune inter-particle interactions. {Fourier transform infrared (}FTIR{) spectroscopy} showed that particle hydrophobicity increases systematically with increase in additive concentration across the {volume phase transition temperature} (VPTT). Zeta potential measurements suggested that the presence of dissociating additives {in the suspension medium} {strongly} suppresses screened {inter-particle} {C}oulombic repulsion{s}. Oscillatory rheological measurements were carried out to understand the impact of {additive-induced modifications in} inter-particle interaction{s} on the viscoelastic properties of dense PNIPAM suspensions. {Reduction in {the effective volume fraction,} $\phi_{eff}$, {due to particle deswelling} {leads} to faster particle-scale dynamics below the VPTT.} {Simultaneously,} we observed that the mechanical moduli values, relaxation timescales, and yield strains exhibited very similar concentration-dependent behaviors in the presence of both dissociating and
non-dissociating additives{, suggesting} that the viscoelastic properties of the suspensions are {governed} {only} by growing particle hydrophobicity.  

{At temperatures a}bove the VPTT{, {the rheology of} samples {prepared} with both dissociating and non-dissociating additives} displayed signatures that are consistent with the formation of colloidal gels. We {show here} that {inter-particle} attractive interaction{s arise} from {the} combined effect of enhanced hydrophobicity and {a} suppression of screened {C}oulombic repulsion {when} dissociating salts {are present} {in the suspension medium}. In the {presence} of sucrose, {however}, inter-particle attraction{s arise} from enhanced particle hydrophobicity {alone}. 
Our cryo-FESEM data was consistent with the rheological results presented earlier and showed relatively denser space-spanning networks with thicker gel strands above the VPTT for samples prepared with dissociating salts.
These observations clearly indicate the increasingly dominant role of electrostatics in driving attractions between the collapsed microgel particles {above the VPTT}. We also observed a two-step yielding behavior {under these conditions} in the presence of additives, {suggesting} {the formation of highly attractive gels}. {T}he relaxation timescale{s} and yield strain values {were seen to} depend strongly on the nature of the additive, again reflecting the importance of electrostatics in driving attractive gel formation above the VPTT. 

The present study shows that the bulk viscoelastic properties of {PNIPAM} microgel suspensions can be {fine-}tuned by introducing suitable additives {even when the microgel concentration remains unchanged}.
The {easy tunability of} interactions between PNIPAM particles can be exploited to control their self-assembly in aqueous suspension. Suspensions of PNIPAM microgels can therefore emerge as excellent candidates in the design of multifunctional materials~\cite{tang2021poly,zheng2015tough}. Fine-tuning the dynamics and rheology of PNIPAM suspensions {\textit{\textit{via}}} the incorporation of additives also holds tremendous promise in {the fabrication of} PNIPAM microgel{-based} actuators~\cite{ansari2022poly}. Additionally, the easy tunability of the VPTT of PNIPAM particles {prepared in additive solutions} makes {microgel systems} a potential candidate for use in various cargo delivery applications at specified temperatures~\cite{ashraf2016snapshot}. {Changing the polarity of the crosslinker {used} during PNIPAM synthesis would provide additional control over the relative contributions of electrostatics and hydrophobicity {in determining material response at temperatures} above the VPTT. }

	
\section*{Conflicts of interest}
There are no conflicts to declare.

\section*{Data availability}
Data will be made available upon reasonable request.

\section*{Acknowledgements}
The authors thank K. M. Yatheendran for assistance in acquiring cryo-FESEM images and K.N. Vasudha for support with DSC and FTIR
experiments.

    \renewcommand\refname{References}
	
	\bibliographystyle{elsarticle-num}
	\bibliography{ref}
    

\section*{Supplementary material}

# ST1 Determination of VPTT via differential scanning calorimetry (DSC):

In these experiments, the effective volume fractions, $\phi_{eff}$, [1] lie between 1.5 in the case of pure water and 0.95 for 0.6M NaCl at 25°C.

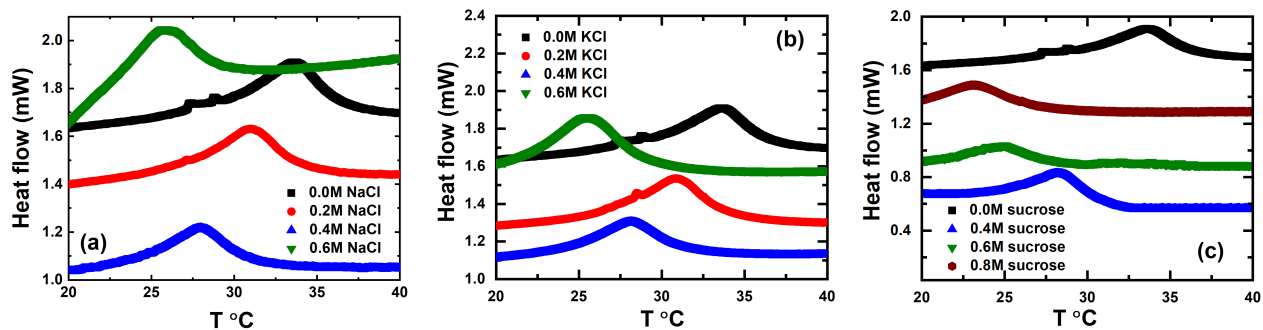

Fig. S1: **Heat flow as a function of temperature, T, for aqueous suspensions of PNIPAM particles prepared with different concentrations of (a) NaCl, (b) KCl, (c) sucrose.**

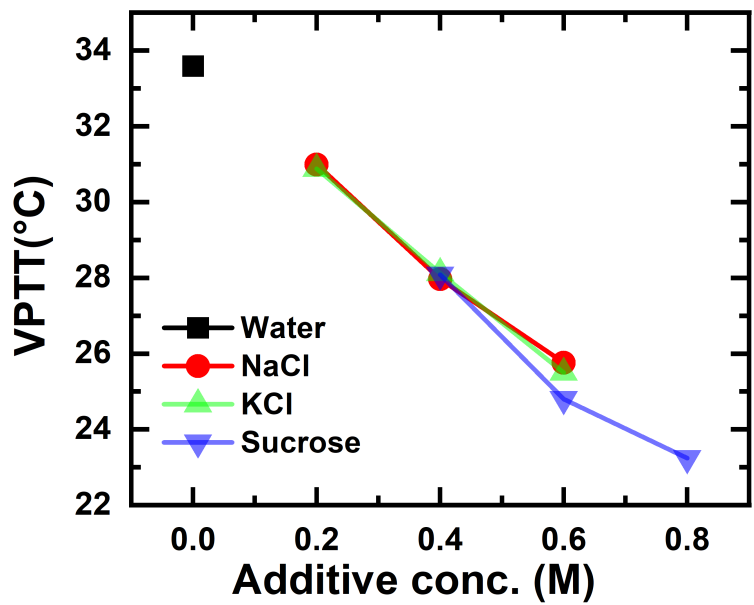

Fig. S2: **Volume phase transition temperature, VPTT, as a function of varying concentrations of dissociating and non-dissociating additives, estimated from DSC data plotted in Fig. S1.**

# ST2 Deconvolution of Fourier transform infrared spectroscopy (FTIR) data to estimate additive induced hydrophobicity:

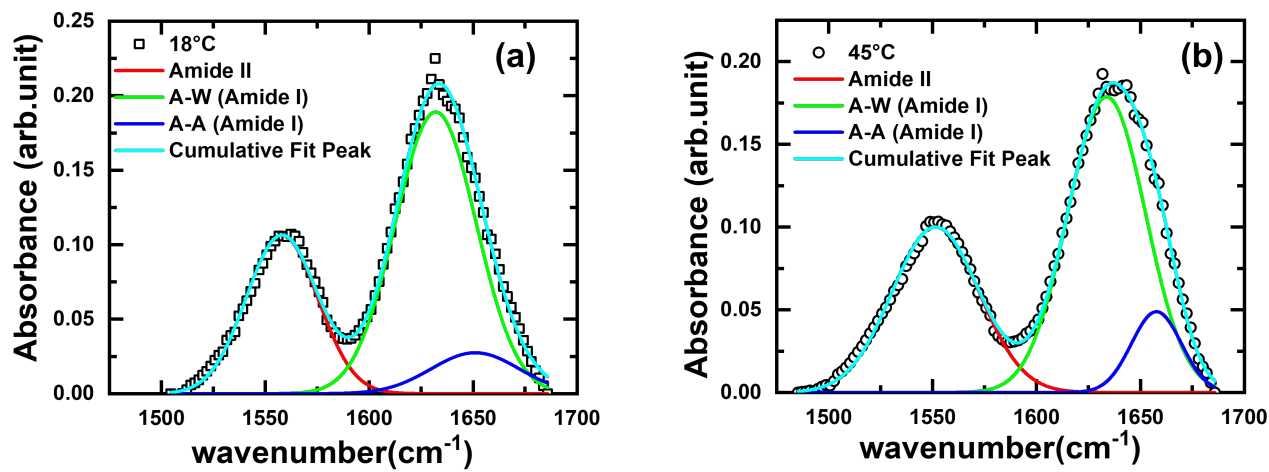

Fig. S3: **FTIR spectra and fits of the amide I and II bands of aqueous suspensions of PNIPAM particles in pure water (a) below the VPTT (18°C) and (b) above the VPTT (45°C). The red curve is the Gaussian fit to the amide II peak at $\sim 1560$ cm$^{-1}$. The amide I peak is deconvoluted into two peaks, wherein the green curve is a Gaussian fit of the peak at $\sim 1625$ cm$^{-1}$ and represents amide-water hydrogen bonding. The blue curve is a Gaussian fit of the peak at $\sim 1652$ cm$^{-1}$ and represents amide-amide hydrogen bonding. The cumulative fit is shown in cyan.**

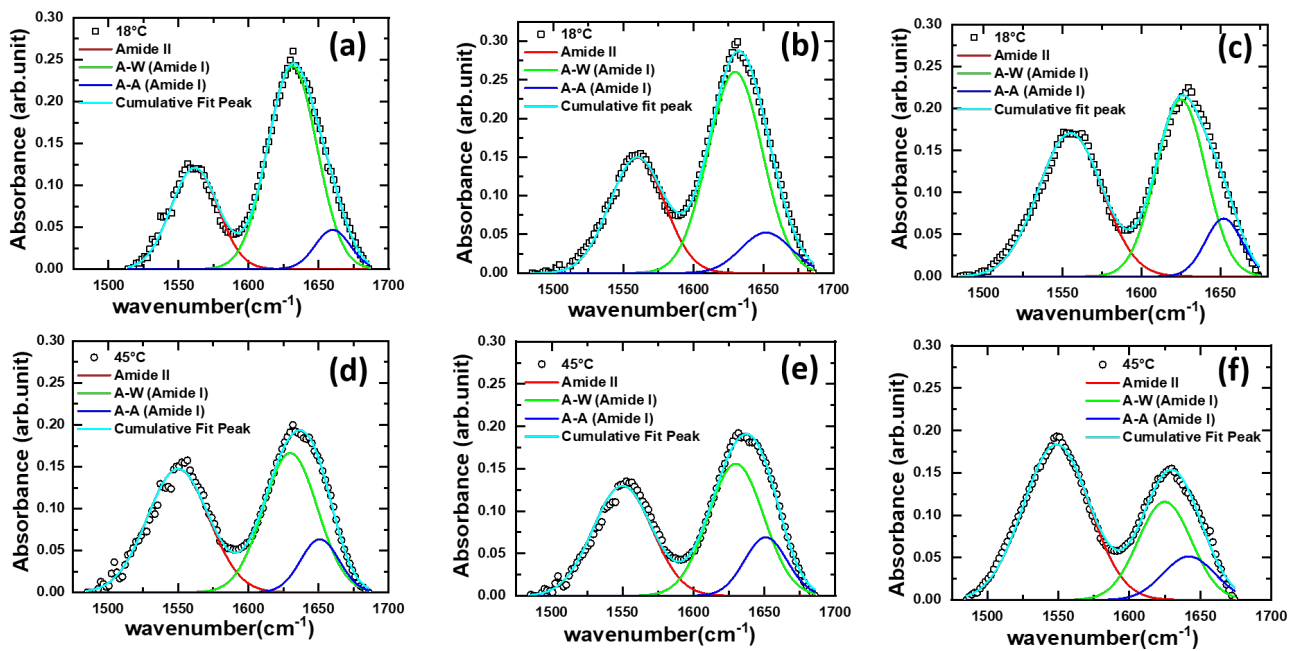

Fig. S4: **(a) FTIR spectra and fits of the amide I and II bands for PNIPAM suspensions with different concentrations of NaCl: (a) 0.2M, (b) 0.4M and (c) 0.6M below the VPTT (18°C). The spectra above the VPTT (45°C) are shown in (d), (e) and (f) respectively. The colours corresponding to the deconvoluted peaks and the sub-figure legends are the same as in Fig. S3.**

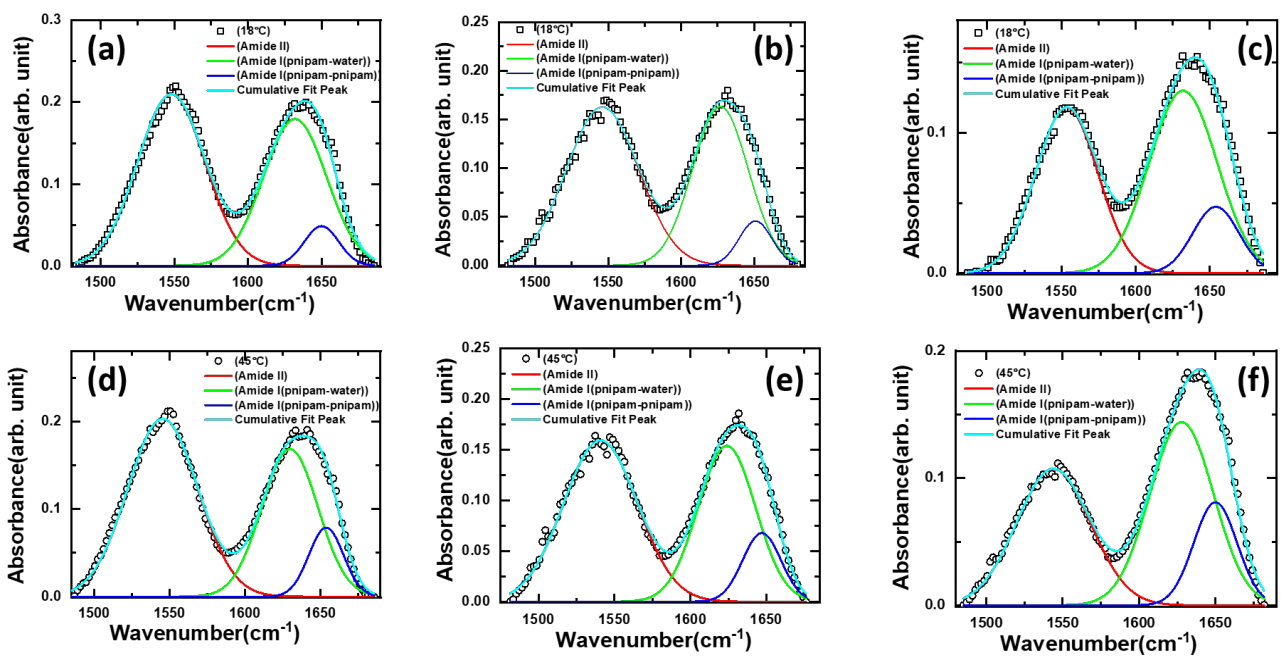

Fig. S5: (a) FTIR spectra and fits of the amide I and II bands for PNIPAM suspensions with different concentrations of KCl: (a) 0.2M, (b) 0.4M and (c) 0.6M below the VPTT (18°C). The spectra above the VPTT (45°C) are shown in (d), (e) and (f) respectively. The colours corresponding to the deconvoluted peaks and the sub-figure legends are the same as in Fig. S3.

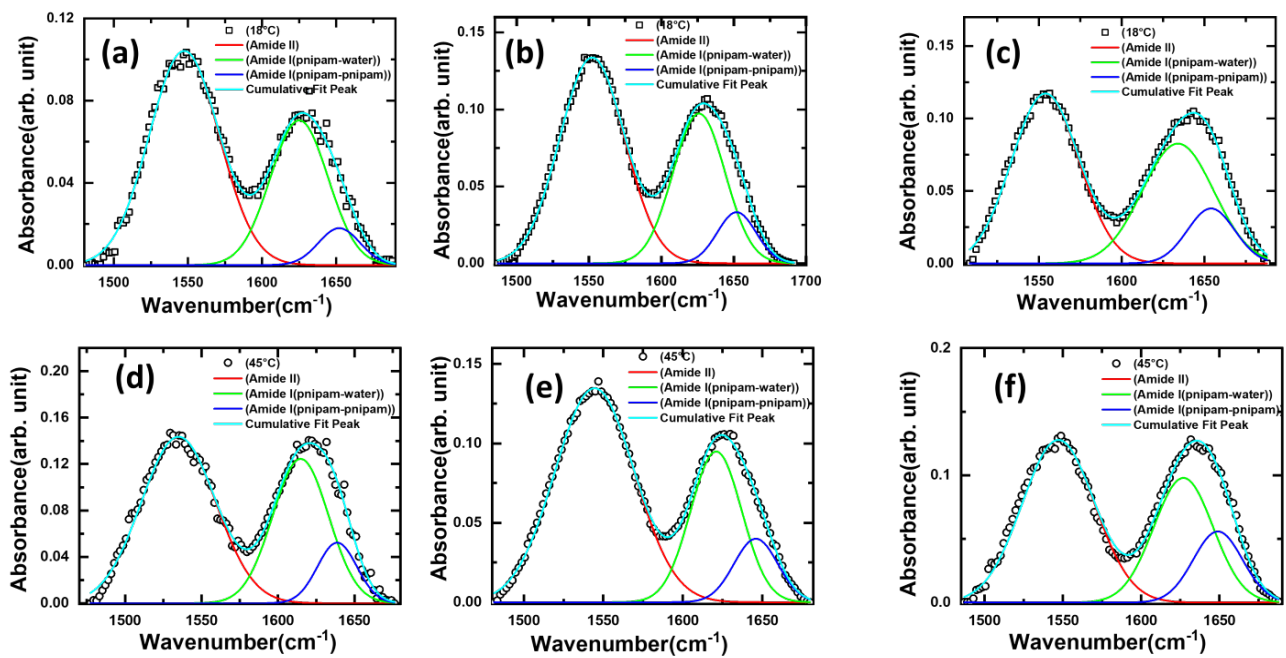

Fig. S6: **(a) FTIR spectra and fits of the amide I and II bands for PNIPAM suspensions with different concentrations of sucrose: (a) 0.4M (b) 0.6M and (c) 0.8M below the VPTT (18°C). The spectra above the VPTT (45°C) are shown in (d), (e) and (f) respectively. The colours corresponding to the deconvoluted peaks and the sub-figure legends are the same as in Fig. S3.**

## ST3 Determination of the hydrodynamic diameter, $<d_H>$, of PNIPAM particles:

Normalized intensity autocorrelation functions, $C(t) = \frac{g^{(2)}(q,t)-1}{A}$, as a function of delay time $t$ were recorded using a Brookhaven Instruments digital autocorrelator (BI 9000T). In the expression for $C(t)$, $A$ is the coherence factor and $g^{(2)}(q,t)$ is defined as $\frac{<I(q,0)I(q,t)>}{<I(q,0)^2>}$, where $I(q,t)$ is the scattered intensity of light at a particular wave vector $q$ and an instant of time $t$. Figs. S7(a-c) show $C(t)$ as a function of $t$ at different temperatures for PNIPAM particles suspended in various additive solutions. For a dilute suspension of polydisperse particles, the temporal decay of $C(t)$ was modeled as $[exp(-(\frac{t}{\tau})^\beta)]^2$. The fitting parameters $\beta$ and $\tau$ were used to calculate the mean relaxation time $<\tau> = (\frac{\tau}{\beta})\Gamma(\frac{1}{\beta})$ [2], where $\Gamma$ is the Euler Gamma function. The mean hydrodynamic diameter, $<d_H>$, of the particles was then estimated using the Stokes-Einstein relationship for spherical particles: $<d_H> = \frac{k_B T <\tau> q^2}{3\pi\eta}$ where $k_B$ is the Boltzmann constant, $T$ is the absolute temperature and $\eta$ is the solvent viscosity.

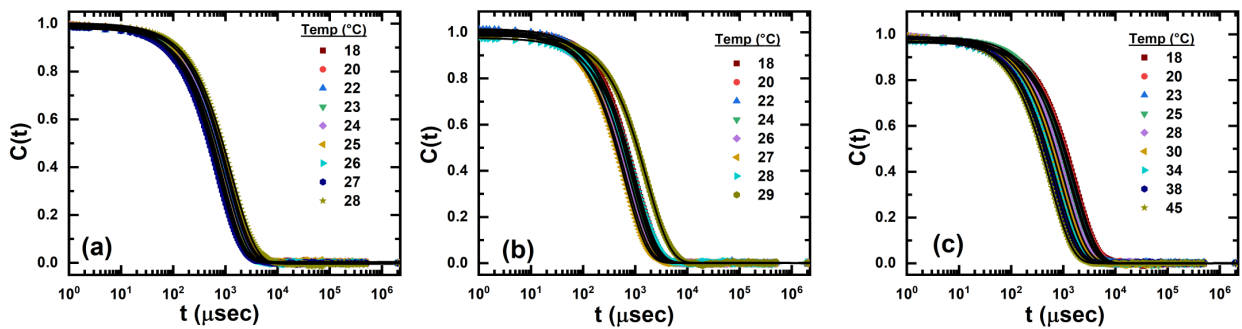

Fig. S7: The normalised intensity autocorrelation functions, $C(t)$, as a function of the delay time, $t$, of PNIPAM particles suspended in 0.6M (a) NaCl, (b) KCl and (c) sucrose solutions. The black solid lines are fits to the expression: $C(t) = [exp(-(\frac{t}{\tau})^\beta)]^2$.

## ST4 Determination of the effective volume fraction, $\phi_{eff}$, of PNIPAM suspensions:

To estimate effective volume fractions, $\phi_{eff}$, aqueous suspensions of PNIPAM particles of known concentrations, c, ranging from 0.01 wt% to 0.9 wt%, were prepared with and without additives. A double gap geometry (DG-26.7) with a gap of 1.886 mm and an effective length of 40 mm was used to measure the viscosity, $\eta$, of these dilute suspensions as a function of strain rate, $\dot{\gamma}_0$, at a constant temperature 25°C. The data was fitted to the Cross model: $\frac{\eta - \eta_\infty}{\eta_0 - \eta_\infty} = \frac{1}{1+(k\dot{\gamma}_0)^m}$, where $\eta_0$ and $\eta_\infty$ correspond to the low and high shear rate viscosity plateaus respectively, $k$ is a time constant related to the relaxation time of the microgel in aqueous suspension, and $m$ is a dimensionless exponent [3]. The inset of Fig. S8 shows representative plots of $\eta$ as a function of $\dot{\gamma}_0$ for PNIPAM suspensions of various concentrations and the corresponding fits to the Cross model.

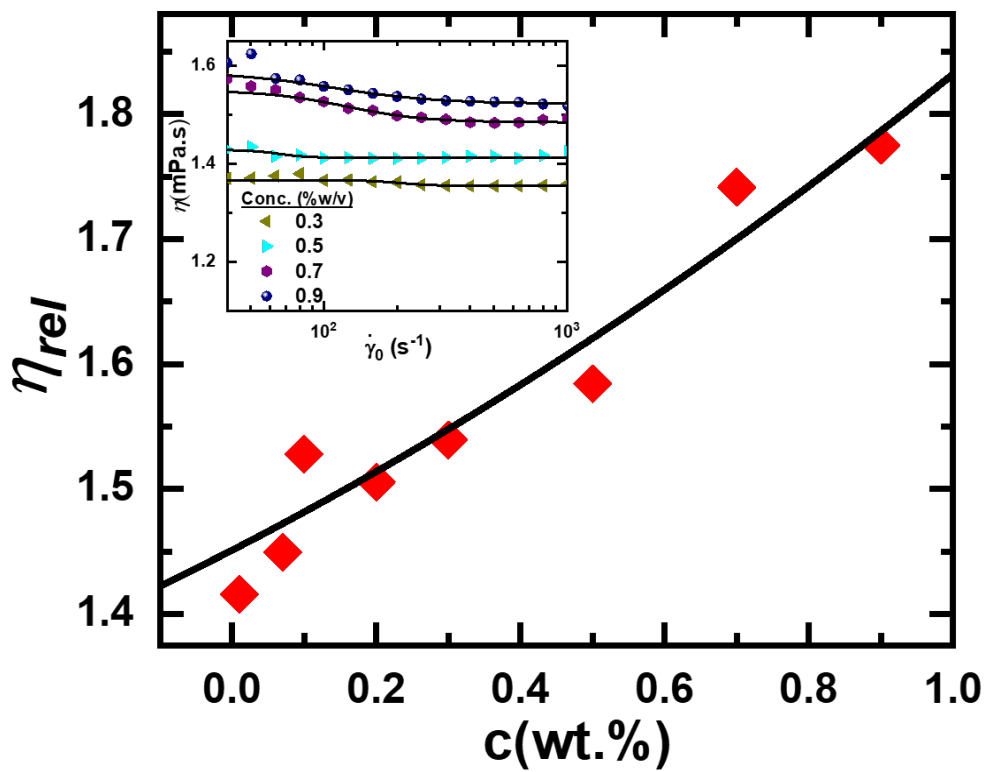

Fig. S8: Relative viscosity, $\eta_{rel}$, as a function of particle concentration, c, of dilute suspensions of PNIPAM particles in pure water. The black line is a fit to the Batchelor's equation. The inset shows the variation of $\eta$ as a function of $\dot{\gamma}_0$ for dilute aqueous suspensions of PNIPAM particles and the fits of the data to the Cross model.

The fitted parameter $\eta_0$ was extracted from fits of the $\eta$ vs $\dot{\gamma}_0$ data to the Cross model, and the relative viscosities, $\eta_{rel}$, of PNIPAM suspensions of various particle concentrations were calculated as $\eta_{rel} = \frac{\eta_0}{\eta_a}$. In this expression, $\eta_a$ is the viscosity of water or additive solution at 25°C. Fig. S8 shows a representative plot of $\eta_{rel}$ as a function of c in pure water and the fit to Batchelor's equation: $\eta_{rel} = 1 + 2.5(\kappa c) + 5.9(\kappa c)^2$ [4, 5]. The fitting parameter, $\kappa$, was extracted from the fits to the $\eta_{rel}$ vs c data. $\kappa$ is a shift factor that converts the mass concentration of PNIPAM to the effective volume fraction. PNIPAM suspensions of desired effective volume fraction were prepared using the relation $\phi_{eff} = \kappa c$. In this study, $\phi_{eff}$ was chosen to be 1.5 at 25°C for PNIPAM suspensions prepared in pure water. The temperature dependent $\phi_{eff}$, was further calculated using the equation $\phi_{eff}(T) = \phi_{eff}(25°C) \times \left[\frac{<d_H(T)>}{<d_H(25°C)>}\right]^3$ [1] and is shown in Fig. S9 for suspensions containing various additives.

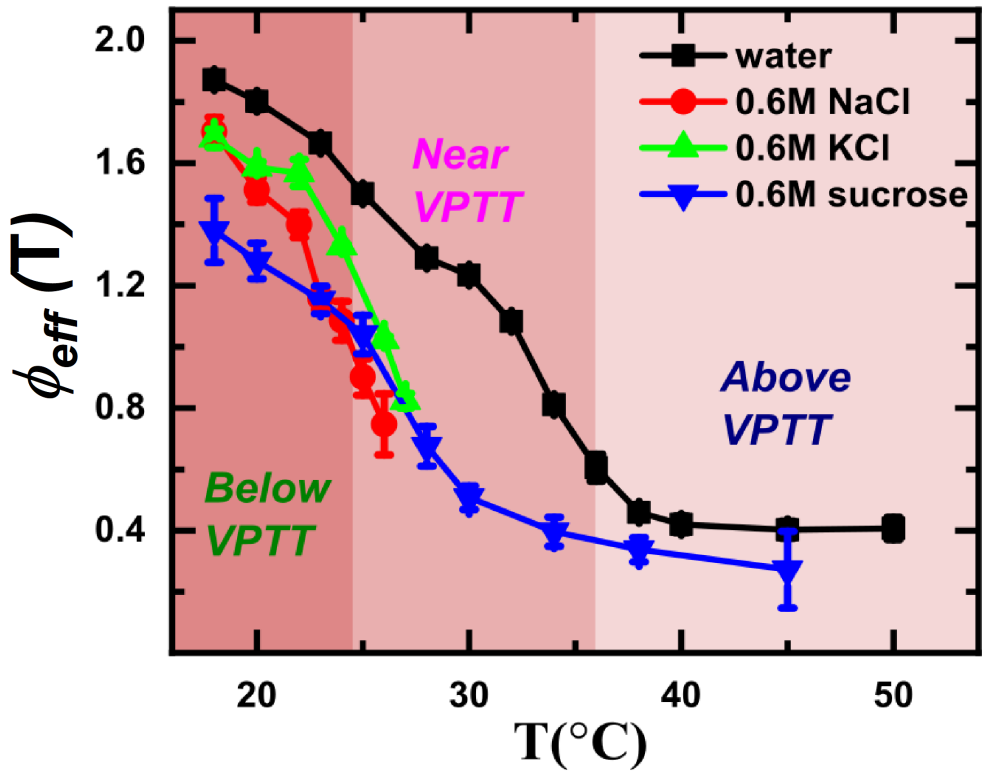

Fig. S9: **Effective volume fraction, $\phi_{eff}$, of PNIPAM particles suspended in water and in various additive solutions.**

# ST5 Temperature-dependent mean hydrodynamic diameters of PNIPAM particles and mechanical moduli of their suspensions:

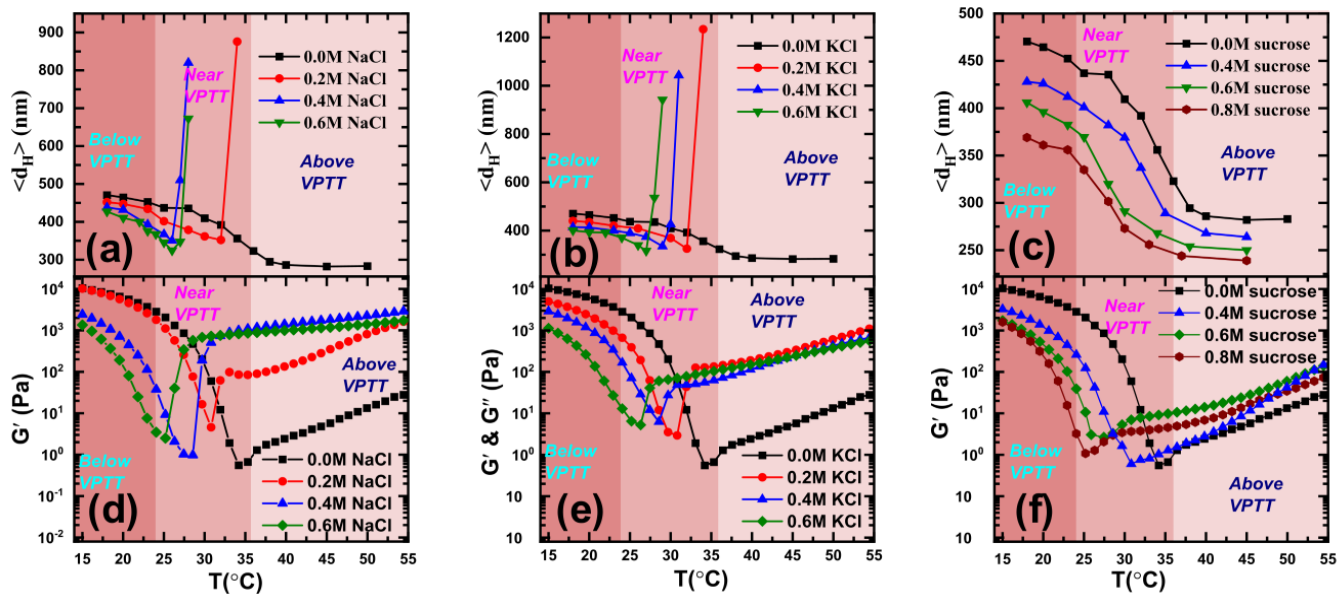

Fig. S10: : (a-c) Temperature-dependent mean hydrodynamic diameters, $<d_H>$, of PNIPAM particles, and (d-f) storage moduli, $G'$, of aqueous suspensions of PNIPAM particles with varying concentrations of (a,d) NaCl, (b,e) KCl and (c,f) sucrose.

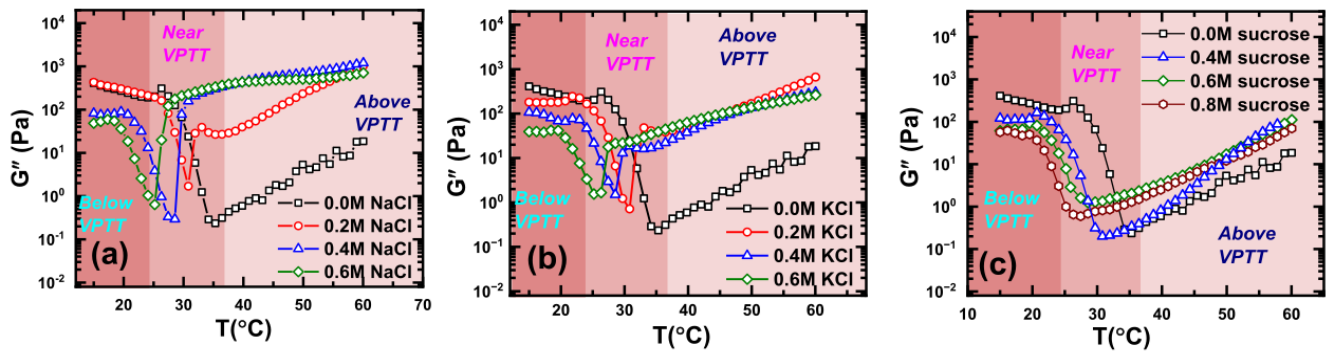

Fig. S11: Temperature-dependent viscous moduli, $G''$, of aqueous suspensions of PNIPAM particles with varying concentrations of (a) NaCl, (b) KCl and (c) sucrose.

## ST6 Frequency sweep & SRFS below the VPTT:

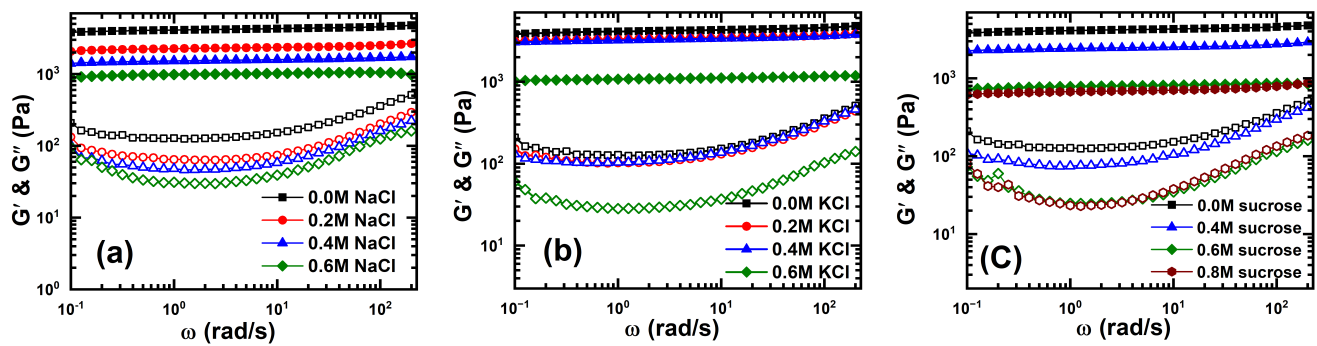

Fig. S12: Elastic moduli, $G'$ (solid symbols), and viscous moduli, $G''$ (hollow symbols), as a function of applied oscillatory frequency, $\omega$, below the VPTT (18°C) for varying concentrations of (a) NaCl, (b) KCl, (c) sucrose.

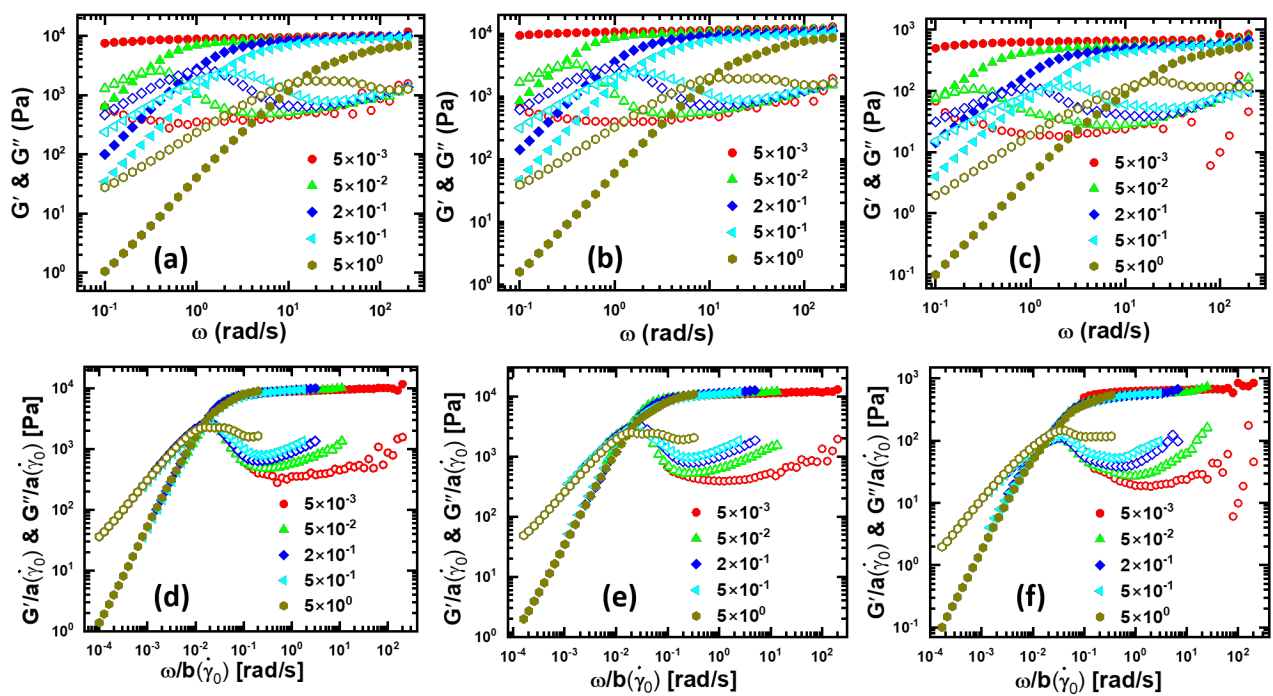

Fig. S13: Elastic moduli, $G'$ (solid symbol), and viscous moduli, $G''$ (hollow symbols), as a function of applied oscillatory frequency, $\omega$, below the VPTT (18°C) for PNIPAM suspensions with (a) 0.2M, (b) 0.4M and (c) 0.6M NaCl measured at five different strain rates, $\dot{\gamma}_0$. Data shown in (a), (b) and (c) are scaled and collapsed on a single master curve and are shown in (d), (e) and (f) respectively.

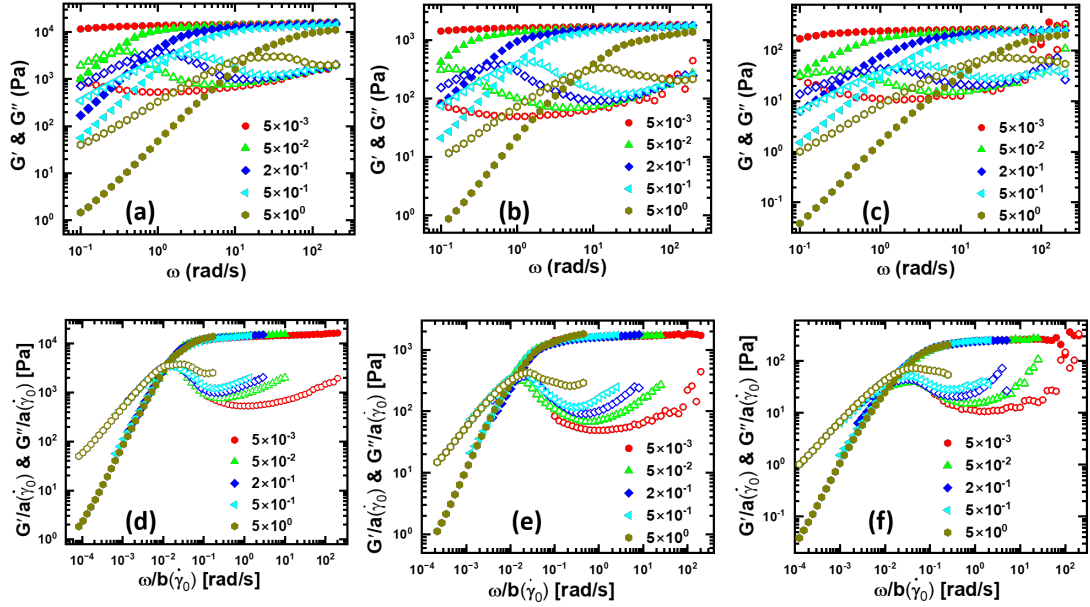

Fig. S14: **Elastic moduli, $G'$ (solid symbol), and viscous moduli, $G''$ (hollow symbols), as a function of applied oscillatory frequency, $\omega$, below the VPTT (18°C) for PNIPAM suspensions with (a) 0.2M, (b) 0.4M and (c) 0.6M KCl measured at five different strain rates, $\dot{\gamma}_0$. Data shown in (a), (b) and (c) are scaled and collapsed on a single master curve and are shown in (d), (e) and (f) respectively.**

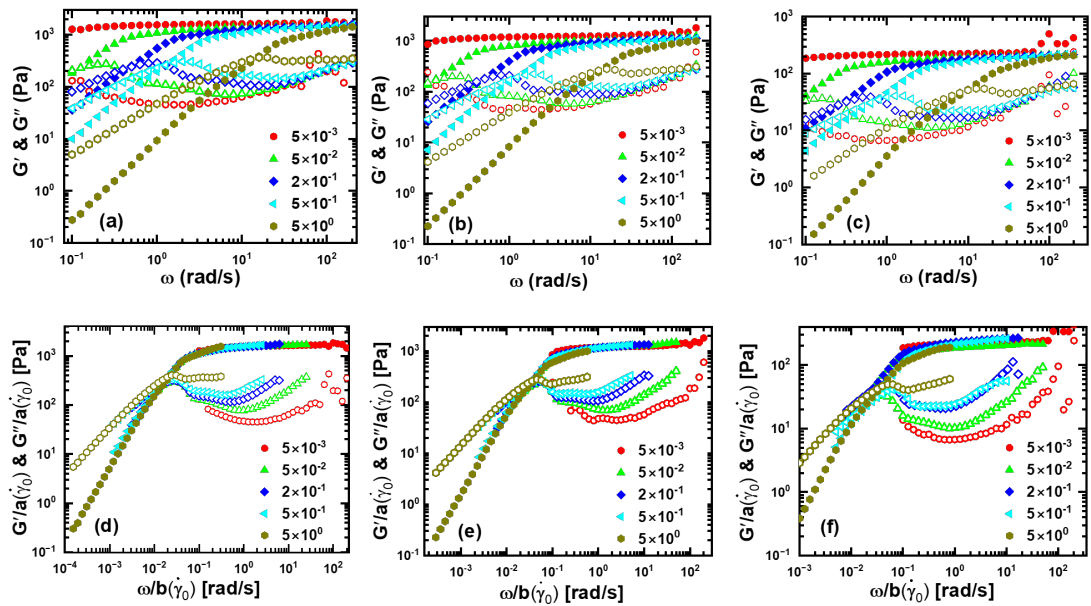

Fig. S15: **Elastic moduli, $G'$ (solid symbol), and viscous moduli, $G''$ (hollow symbols), as a function of applied oscillatory frequency, $\omega$, below the VPTT (18°C) for PNIPAM suspensions with (a) 0.4M, (b) 0.6M and (c) 0.8M sucrose measured at five different strain rates, $\dot{\gamma}_0$. Data shown in (a), (b) and (c) are scaled and collapsed on a single master curve and are shown in (d), (e) and (f), respectively.**

## ST7 Amplitude sweep below the VPTT:

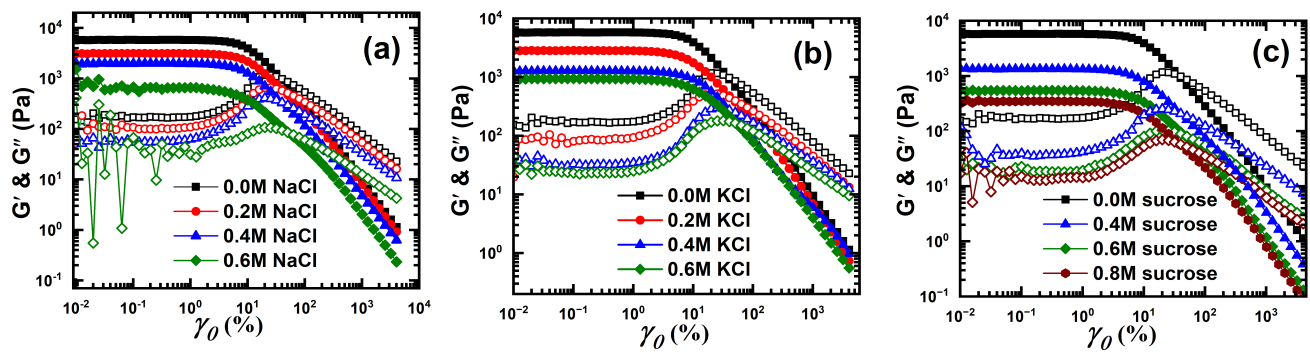

Fig. S16: **: Elastic moduli, $G'$ (solid symbol), and viscous moduli, $G''$ (hollow symbols), as a function of applied oscillatory amplitudes, $\gamma_0$, below the VPTT (18°C) with varying concentrations of (a) NaCl, (b) KCl, (c) sucrose.**

## ST8 Estimation of yield strain below the VPTT:

Fig. S17 shows a representative plot of the variation of oscillatory stress amplitude, $\sigma_0$, as a function of applied strain amplitude, $\gamma_0$, below the VPTT for a dense suspension of PNIPAM particles in 0.4M KCl solution. At low strains, $\sigma_0$ is found to vary linearly with $\gamma_0$. Following a previous work [6] and as shown using a green dashed line in Fig. S17, we identified the value of $\gamma_0$ at which $\sigma_0$ starts to deviate by more than 3% from the linear regime as the yield strain, $\gamma_y$.

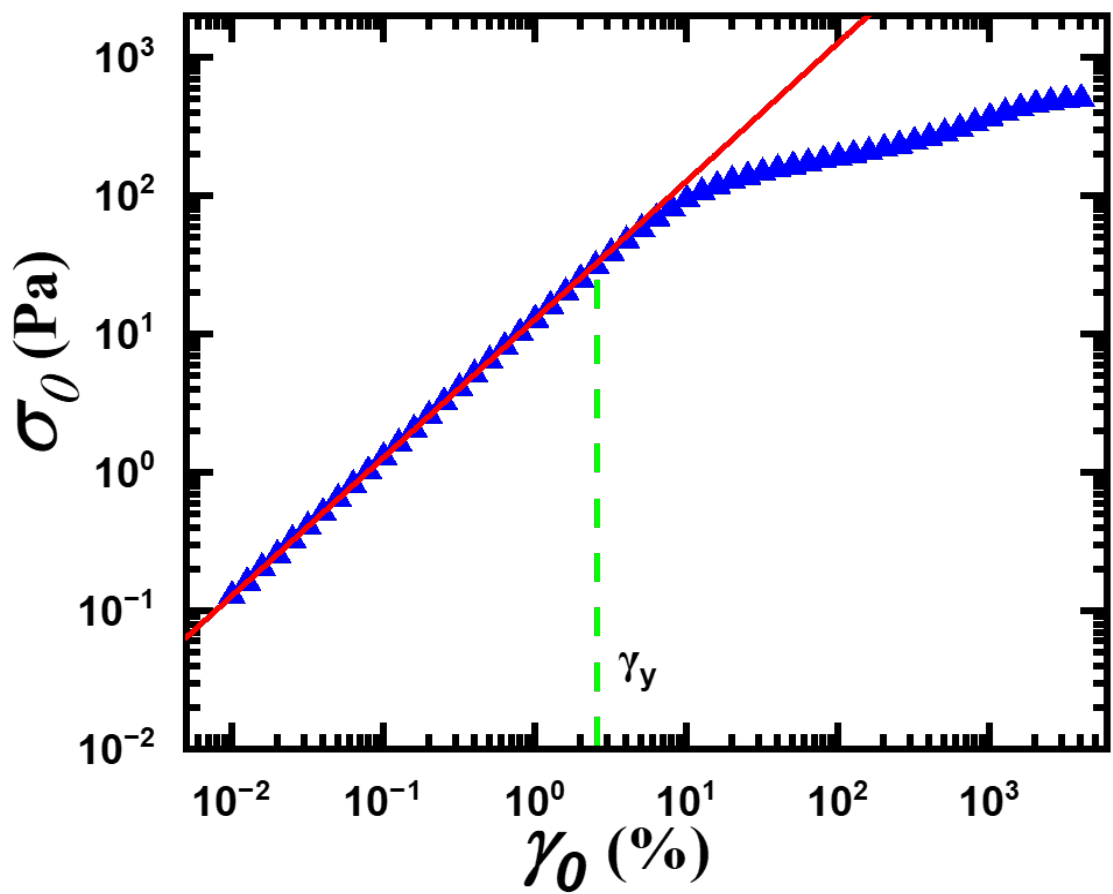

Fig. S17: **Stress amplitude, $\sigma_0$, vs. oscillatory strain amplitude, $\gamma_0$, below the VPTT (18°C) for a dense suspension of PNIPAM particles with 0.4M KCl. The red line is the linear fit to the data in the small-strain regime. The vertical green dashed line indicates the strain value at which there is a significant deviation from Hooke's law. This strain value is identified as the yield strain $\gamma_y$.**

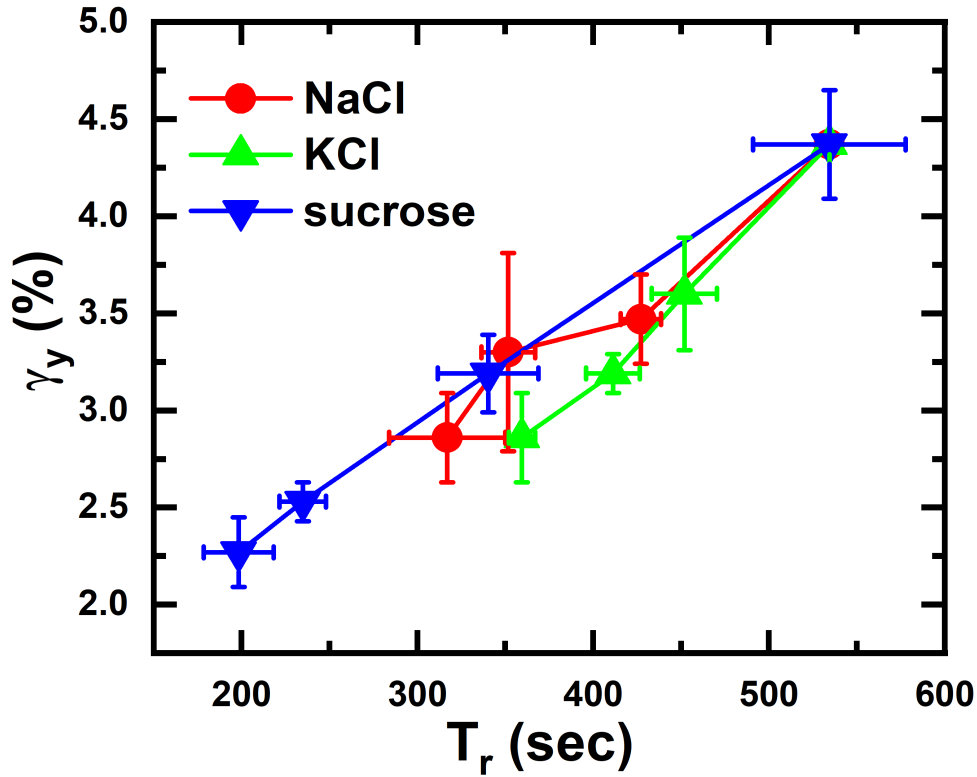

Fig. S18: **Variation of yield strain, $\gamma_y$, as a function of structural relaxation timescale, $T_r$, below the VPTT.**

## ST9 Frequency sweep above the VPTT:

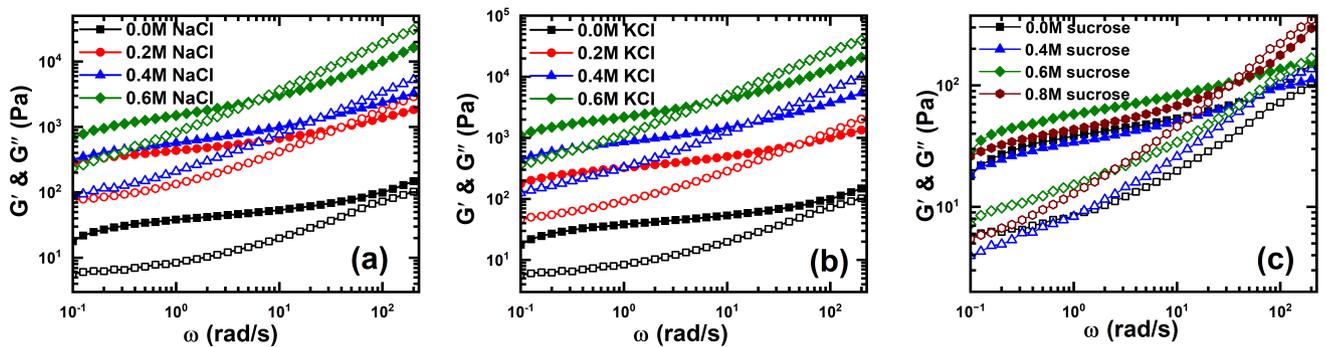

Fig. S19: **Elastic moduli, $G'$ (solid symbol), and viscous moduli, $G''$ (hollow symbols), as a function of applied oscillatory frequency, $\omega$, above the VPTT (45°C) with varying concentrations of (a) NaCl, (b) KCl, (c) sucrose.**

## ST10 Amplitude sweep above the VPTT:

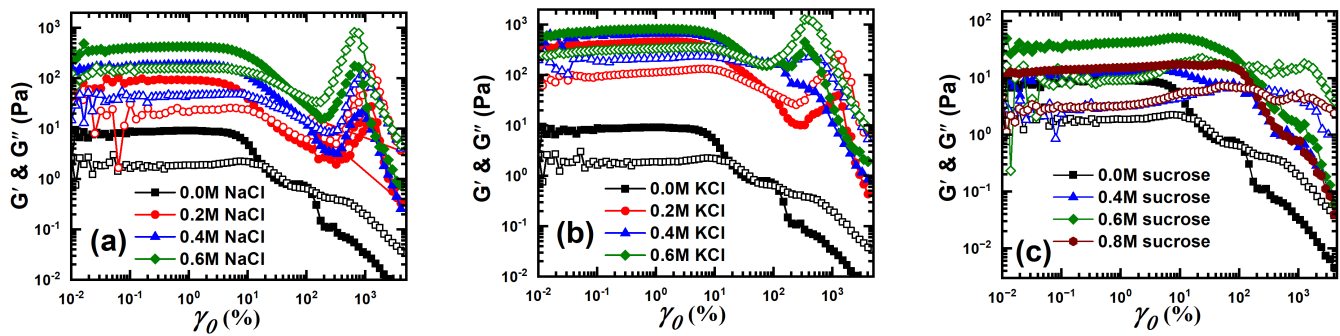

Fig. S20: Elastic moduli, $G'$ (solid symbol), and viscous moduli, $G''$ (hollow symbols), as a function of applied oscillatory strain amplitudes, $\gamma_0$, above the VPTT (45°C) with varying concentrations of (a) NaCl, (b) KCl, (c) sucrose.

## ST11  Estimation of yield strain above the VPTT:

Fig. S21 shows a representative plot of the variation of oscillatory stress amplitude, $\sigma_0$, as a function of applied strain amplitude, $\gamma_0$, above the VPTT of a dense suspension of PNIPAM particles in 0.4M KCl solution. We observe that the suspension shows a double-yielding phenomenon above the VPTT. The first yield strain, $\gamma_{y1}$, is measured from the deviation of the stress response from linearity as discussed in ST8. The second yield strain, $\gamma_{y2}$, is estimated by directly considering the $\gamma_0$ at which the $\sigma_0$ peaks. The first and the second yield strains are indicated by dashed green and red lines respectively in Fig. S21.

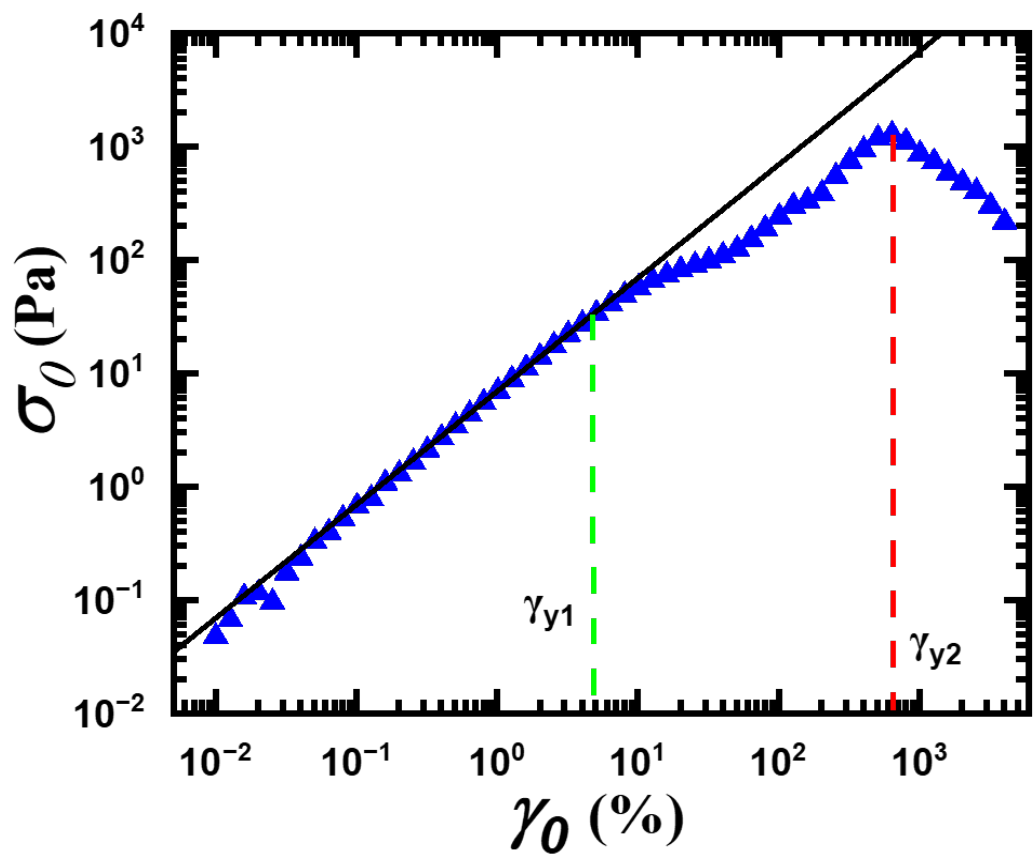

Fig. S21: Stress amplitude, $\sigma_0$, vs. oscillatory strain amplitude, $\gamma_0$, above the VPTT (45°C) for a dense suspension of PNIPAM particles with 0.4M KCl. The black line is the fit to the data in the linear regime. Vertical dashed lines display the first and second yield strain values.

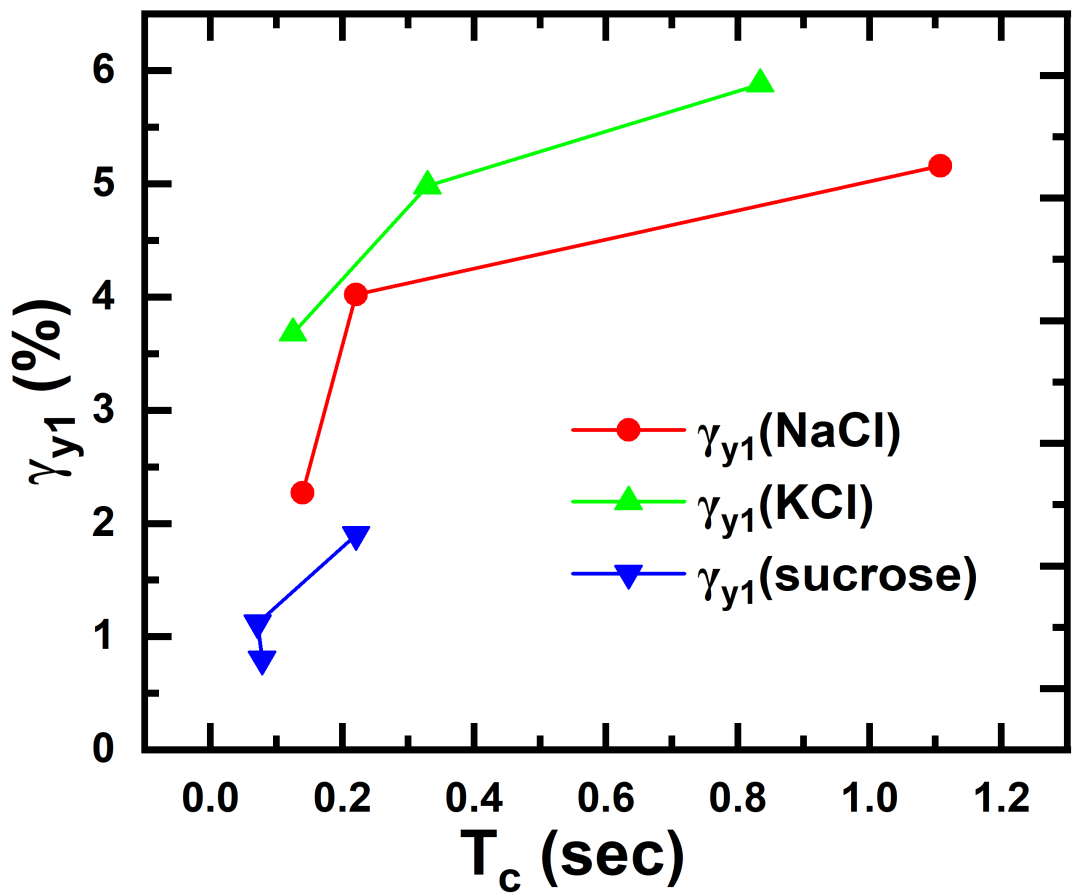

Fig. S22: **Variation of first yield strain, $\gamma_{y1}$, as a function of crossover timescale, $T_c$, above the VPTT.**


\end{document}